\documentclass[aps,prl,reprint,superscriptaddress]{revtex4-2}

\usepackage{float}
\usepackage{stmaryrd}
\usepackage{amsmath} 

\usepackage{gensymb}
\usepackage{textcomp}
\usepackage{array}

\usepackage{graphicx}
\usepackage{amsmath}
\usepackage{amssymb}
\usepackage{epstopdf}
\usepackage{bm}
\usepackage[colorinlistoftodos]{todonotes}
\usepackage{siunitx}
\usepackage[absolute,overlay]{textpos}

\usepackage{upgreek}
\usepackage[colorlinks=true, citecolor={black}, urlcolor={blue!60!black}, linkcolor = {black}]{hyperref}
\usepackage[nameinlink]{cleveref}
\Crefname{figure}{Fig.}{Figs.}

\usepackage{comment}
\usepackage{bold-extra}
\usepackage{soul}

\makeatletter
\newcommand{\fmarki}{*}
\newcommand{\fmarkii}{\ensuremath{\dagger}}
\newcommand{\fmarkiii}{\ensuremath{\ddagger}}
\newcommand{\fmarkiv}{\ensuremath{\mathsection}}
\newcommand{\fmarkv}{\ensuremath{\mathparagraph}}
\newcommand{\fmarkvi}{\ensuremath{\|}}
\newcommand{\fmarkvii}{**}
\newcommand{\fmarkviii}{\ensuremath{\dagger\dagger}}
\newcommand{\fmarkix}{\ensuremath{\ddagger\ddagger}}

\newcommand{\bpm}{\begin{pmatrix}}
\newcommand{\epm}{\end{pmatrix}}

\def\@fnsymbol#1{{\ifcase#1\or \fmarki\or \fmarkii\or \fmarkiii\or \fmarkiv\or \fmarkv\or \fmarkvi\or \fmarkvii\or \fmarkviii\or \fmarkix \else\@ctrerr\fi}}
\makeatother

\renewcommand{\fmarki}{\ensuremath{\dagger}}
\renewcommand{\fmarkii}{*}

\def\loc {$G_{\mathrm{l}}$}
\def\nloc {$G_{\mathrm{nl}}$}
\def\gammaO {$\Gamma_{\mathrm{o}}$}
\def\gammaE {$\Gamma_{\mathrm{e}}$}

\begin{document}

\title{A two-site Kitaev chain in a two-dimensional electron gas}

\author{Sebastiaan L.D. ten Haaf}
\altaffiliation{These authors contributed equally to this work.}
\affiliation{QuTech and Kavli Institute of Nanoscience, Delft University of Technology, Delft, 2600 GA, The Netherlands}

\author{Qingzhen Wang}
\altaffiliation{These authors contributed equally to this work.}
\affiliation{QuTech and Kavli Institute of Nanoscience, Delft University of Technology, Delft, 2600 GA, The Netherlands}

\author{A. Mert Bozkurt}
\affiliation{QuTech and Kavli Institute of Nanoscience, Delft University of Technology, Delft, 2600 GA, The Netherlands}

\author{Chun-Xiao Liu}
\affiliation{QuTech and Kavli Institute of Nanoscience, Delft University of Technology, Delft, 2600 GA, The Netherlands}

\author{Ivan Kulesh}
\affiliation{QuTech and Kavli Institute of Nanoscience, Delft University of Technology, Delft, 2600 GA, The Netherlands}

\author{Philip Kim}
\affiliation{QuTech and Kavli Institute of Nanoscience, Delft University of Technology, Delft, 2600 GA, The Netherlands}

\author{Di~Xiao}
\affiliation{Department of Physics and Astronomy, Purdue University, West Lafayette, 47907, Indiana, USA}

\author{Candice Thomas}
\affiliation{Department of Physics and Astronomy, Purdue University, West Lafayette, 47907, Indiana, USA}

\author{Michael J. Manfra}
\affiliation{Department of Physics and Astronomy, Purdue University, West Lafayette, 47907, Indiana, USA}
\affiliation{Elmore School of Electrical and Computer Engineering, ~Purdue University, West Lafayette, 47907, Indiana, USA}
\affiliation{School of Materials Engineering, Purdue University, West Lafayette, 47907, Indiana, USA}

\author{Tom Dvir}
\affiliation{QuTech and Kavli Institute of Nanoscience, Delft University of Technology, Delft, 2600 GA, The Netherlands}

\author{Michael Wimmer}
\affiliation{QuTech and Kavli Institute of Nanoscience, Delft University of Technology, Delft, 2600 GA, The Netherlands}

\author{Srijit Goswami}\email{s.goswami@tudelft.nl}
\affiliation{QuTech and Kavli Institute of Nanoscience, Delft University of Technology, Delft, 2600 GA, The Netherlands}

\begin{abstract}

Artificial Kitaev chains can be used to engineer Majorana bound states (MBSs) in superconductor-semiconductor hybrids~\cite{DasSarma2012, Leijnse2012, Fulga2013,Dvir2023}.
In this work, we realize a two-site Kitaev chain in a two-dimensional electron gas by coupling two quantum dots through a region proximitized by a superconductor. 
We demonstrate systematic control over inter-dot couplings through in-plane rotations of the magnetic field and via electrostatic gating of the proximitized region.
This allows us to tune the system to sweet spots in parameter space, where robust correlated zero bias conductance peaks are observed in tunnelling spectroscopy. 
To study the extent of hybridization between localized MBSs, we probe the evolution of the energy spectrum with magnetic field and estimate the Majorana polarization, an important metric for Majorana-based qubits~\cite{Tsintzis2022,tsintzis2023}. 
The implementation of a Kitaev chain on a scalable and flexible 2D platform provides a realistic path towards more advanced experiments that require manipulation and readout of multiple MBSs. 
 
\end{abstract}

\maketitle
\section{Introduction}
Superconductor-semiconductor hybrid systems have been intensively investigated as a potential platform to engineer topologically protected Majorana bound states (MBSs). 
In particular, significant efforts have been dedicated to studying one-dimensional systems coupled to s-wave superconductors~\cite{Lutchyn_PRL_2010, Oreg_PRL_2010,Prada2020}. 
However, uncontrolled microscopic variations in hybrid devices have complicated the study of MBSs~\cite{CX_ABS_MBS_2017,Vuik_quasi_Majorana_scipost_2019,Pan_Physical_mechanism_PRR_2020}.
A potential way to mitigate the effects of disorder, is to create a Kitaev chain~\cite{Kitaev2001} using an array of quantum dots~(QDs) with controllable couplings~\cite{DasSarma2012, Leijnse2012, Fulga2013}. 
In fact, a chain consisting of only two~QDs, while not topologically protected, is sufficient to create localized MBSs~\cite{Leijnse2012}. 
These so-called ``poor man's Majoranas" have recently been realized in nanowires~\cite{Dvir2023}, which has led to proposals~\cite{Gabor2020, Boross2023, Liu2023b, tsintzis2023} to study non-Abelian statistics by fusing or braiding MBSs in multiple two-site chains.
However, in order to perform these studies, and move towards a Majorana-based qubit with integrated readout and control, it is vital to have a scalable and flexible 2D architecture. 

In this work we realize a two-site Kitaev chain by coupling two spin-polarized QDs in an InSbAs two-dimensional electron gas (2DEG). 
By tuning the couplings between the QDs to so-called ``sweet spots", we demonstrate correlated zero-bias conductance peaks (ZBPs) that are resilient to local perturbations. 
In addition to electrostatic control, we show that the planar 2DEG geometry allows one to reach such sweet spots through an in-plane rotation of the magnetic field. 
An important prerequisite to produce  localized MBSs, is that the Zeeman splitting in the QDs is sufficiently large~\cite{Tsintzis2022}.
Surprisingly, however, we find that several features used to identify ``sweet spots" (such as correlated ZBPs) actually persist down to zero magnetic field.
We show that the evolution of the energy spectrum with magnetic field provides complementary information, which allows us to estimate the Majorana polarization, a metric quantifying the extent of hybridization between MBSs \cite{Bena2015, Tsintzis2022,tsintzis2023}. 
    
\section*{Results}
\subsection*{Model with strongly coupled dots}
\begin{figure*}[t!]
\centering
\includegraphics[width=\textwidth]{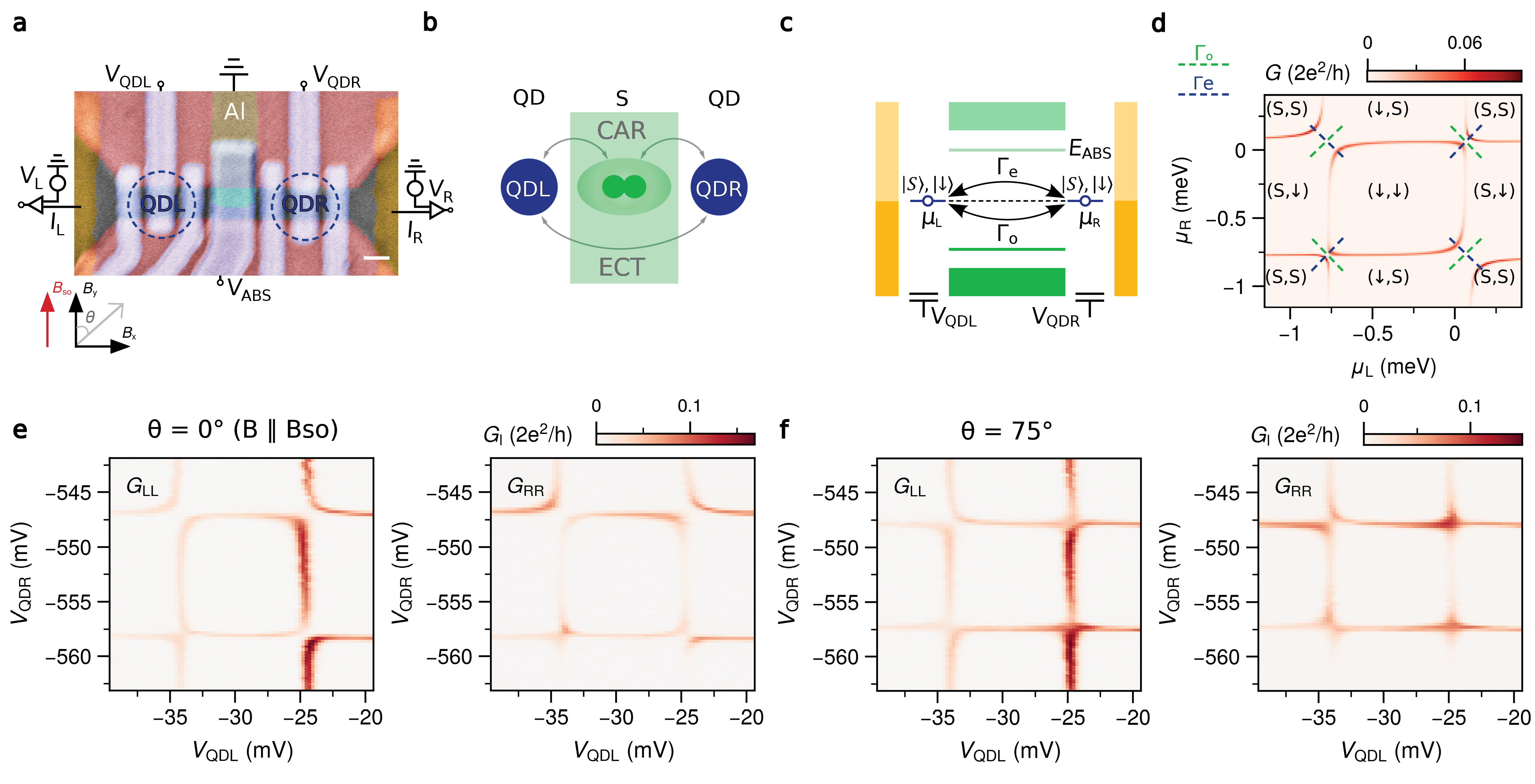}
\caption{\textbf{Device, model and CSDs}
\textbf{(a)} False-colored scanning electron micrograph of device A.
Inset axis shows coordinates of the external magnetic field and the expected spin-orbit field direction w.r.t. the device orientation. 
The gate-defined QDs (QDL and QDR) are indicated.
Scale bar is \SI{100}{nm}.  
\textbf{(b)} Sub-gap transport processes between the QDs and the superconductor. ECT exchanges an electron between the QDs, while CAR allows for pairwise exchange of two electrons with the superconductor (SC). 
\textbf{(c)} Energy level diagram showing 
two spin-polarized YSR-states in two QDs that are coupled through a hybrid section.
An Andreev bound state (at energy $E_{\mathrm{ABS}}$) mediates two types of virtual tunnel couplings between the QDs, denoted by \gammaE~and \gammaO.
\textbf{(d)} Numerically calculated CSD of two coupled QDs, in the absence of spin-orbit coupling. Dashed lines indicate which states are expected to be hybridized through each type of coupling.
\textbf{(e)} Measured CSD across two charge degeneracy points in QDL and QDR with $\theta = 0\degree$ ($B\parallel B_{\mathrm{SO}}$), corresponding to numerical conductance in (d).
\textbf{(f)} Measured CSD with $\theta = 75\degree$. Data in (e) and (f) are taken at $V_{\mathrm{ABS}}$ =\SI{-624}{mV} and $B = \SI{100}{\mathrm{mT}}$.
}
\label{fig1}
\end{figure*}
The Kitaev model~\cite{Kitaev2001} can be implemented by coupling spin-polarized QDs via Andreev bound states (ABSs) in a semiconductor-superconductor hybrid~\cite{Dvir2023}. 
Coupling between the QDs is mediated by two types of coherent tunneling processes, as illustrated in \Cref{fig1}b. 
A hopping interaction arises through elastic co-tunnelling (ECT) and a pairing interaction arises via the creation or breaking of a Cooper pair in the superconductor through crossed Andreev reflection (CAR). 
To emulate a Kitaev chain, the relative amplitudes of these processes must be controlled~\cite{Liu2022,Bordin2022,Wang2022a}.
Furthermore, large inter-dot couplings are desired in order to isolate zero-energy MBSs from higher-energy excitations~\cite{Leijnse2012}.
This can be achieved by increasing tunnelling rates between the QDs and the proximitized region, additionally inducing superconducting correlations in the QDs~\cite{Liu2023, Zatelli2023}.
In this regime the QDs can be described as Yu-Shiba-Rusinov (YSR) states~\cite{Meng2009,Rasmussen2009,Deacon2010, Lee2014,Jellinggaard2016,Rasmussen2018}.

An energy level diagram of the system is shown in \Cref{fig1}c, where, at finite magnetic field, the ground state of each proximitized QD is either a doublet state $\lvert \downarrow \rangle$ or a singlet state $\lvert \mathrm{S}\rangle$.
The electro-chemical potential of the QDs are denoted $\upmu_{\mathrm{L}}$ and $\upmu_{\mathrm{R}}$.
We consider the combined state of the QDs $\rvert \sigma_{\mathrm{L}}$,$\sigma_{\mathrm{R}}\rangle$, where $\sigma_{\mathrm{L}},\sigma_{\mathrm{R}}$ $\in$ ($\lvert \mathrm{S}\rangle$, $\lvert \downarrow \rangle$).
In this description, ECT and CAR processes give rise to two types of effective couplings. 
States with total odd-parity ($\vert \mathrm{S}, \downarrow \rangle$ and $\vert$$\downarrow, \mathrm{S} \rangle$) have the same total spin ($\frac{1}{2}$) and therefore couple through a spin-conserving term \gammaO. States with total even-parity ($\vert \mathrm{S}, \mathrm{S}\rangle$ and $\vert$$\downarrow ,\downarrow \rangle$) have different total spin (0 or 1) and couple through a spin non-conserving term \gammaE.
Similar to a system with non-proximitized QDs~\cite{Leijnse2012,DasSarma2012}, MBSs should arise when these couplings are equal (\gammaO$=$\gammaE)~\cite{Liu2023}, as further described in SI-Methods. 
\Cref{fig1}d shows the numerically obtained conductance $G$, considering the local transport, as a function of $\upmu_{\mathrm{L}}$ and $\upmu_{\mathrm{R}}$ (details of the model can be found in SI-Methods).
This charge stability diagram (CSD) reveals avoided crossings at the charge degeneracy points, indicative of strong inter-dot coupling.
In the absence of spin-orbit interaction only the spin-conserving coupling \gammaO~is relevant, thus strongly hybridizing the odd-parity states. 
Horizontal and vertical conductance features are visible between avoided crossings as a result of local Andreev reflection, typical for YSR-states. \newline

\subsection{Device description}
A scanning electron micrograph of a typical device (device A) is shown in \Cref{fig1}a. Gate-defined QDs are created on the left and right of a region proximitized by aluminium (green).
The QDs are strongly coupled to the superconductor, resulting in the formation of sub-gap YSR states detailed in Fig.~S1).
Biases applied to the left and right leads  ($V_{\mathrm{L}}$ and $V_{\mathrm{R}}$) can be varied and the currents in the left and right leads ($I_{\mathrm{L}}$ and $I_{\mathrm{R}}$) can be measured simultaneously.
Using standard lock-in techniques, we measure local conductances $G_{\mathrm{LL}}$ $  (\frac{\mathrm{d}I_\mathrm{L}}{\mathrm{d}V_\mathrm{L}}$) and $G_{\mathrm{RR}}$ $ (\frac{\mathrm{d}I_\mathrm{R}}{\mathrm{d}V_\mathrm{R}})$, denoted $G_{\mathrm{l}}$, and non-local conductances $G_{\mathrm{LR}}$ ($ \frac{\mathrm{d}I_\mathrm{L}}{\mathrm{d}V_\mathrm{R}}$) and $G_{\mathrm{RL}}$ ($\frac{\mathrm{d}I_\mathrm{R}}{\mathrm{d}V_\mathrm{L}}$), denoted $G_{\mathrm{nl}}$.
We report on two similar devices. Device A was used for measurements in Fig.~1, Fig.~3 and Fig.~4. Device B (image shown in S2a) was used to obtain the measurements in Fig. 2. All measurements are performed in a dilutionrefrigerator, with a base temperature of \SI{20}{mK}. 

By applying a magnetic field along the spin-orbit field, the effect of the spin-orbit interaction is suppressed (see Fig.~S2), as previously observed in similar devices~\cite{WangQ2022}.
We measure $G_{\mathrm{l}}$, as $V_{\mathrm{QDL}}$ and $V_{\mathrm{QDR}}$ are swept across two charge degeneracy points in each QD, resulting in the CSDs shown in \Cref{fig1}e. 
Similar to the simulations (\Cref{fig1}d), we find avoided crossings that indicate a strong coupling between odd parity states, i.e.,~ \gammaO$>$\gammaE.
Next, we rotate the external magnetic field away from the spin-orbit field, allowing spin non-conserving processes to occur. 
This is reflected in the avoided crossings in the CSDs (\Cref{fig1}f), where we indeed see that even-parity states can now hybridize, indicating a sizeable \gammaE.
In particular, the top left and bottom left avoided crossings have changed direction, indicating an even parity ground state at these charge degeneracy points.
The evolution from \Cref{fig1}e to \Cref{fig1}f suggests that the field angle can be used to tune the system into the sweet spot  (\gammaO=\gammaE) where MBSs emerge. \\

\subsection*{Tuning to the Majorana sweet spot}
\begin{figure}[t!]
\centering
\includegraphics[width = 0.49\textwidth]{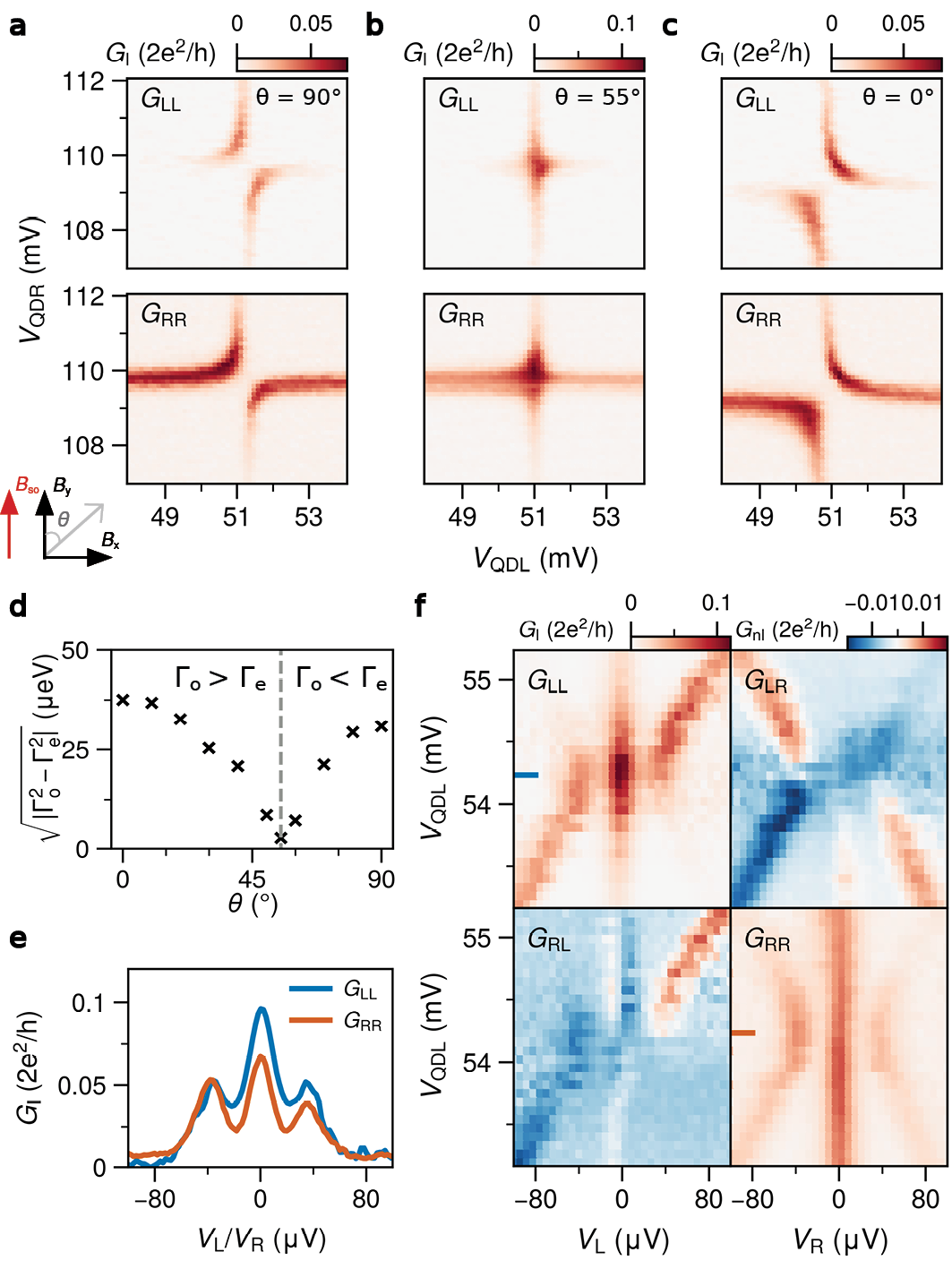}
\caption{\textbf{Tuning \gammaE~and \gammaO~with magnetic field angle.}  
Measurements obtained for device B, characterized in Fig.~S2.  
\textbf{(a)}~  CSD showing a diagonal avoided crossing (\gammaE $ > $\gammaO), obtained with $B\perp B_{\mathrm{SO}}$. 
\textbf{(c)} Rotating the field such that $B\parallel B_{\mathrm{SO}}$, the avoided crossing changes direction (\gammaE$ \approx <$\gammaO).
\textbf{(b)} At $\SI{55}{\degree}$ the the avoided crossing disappears (\gammaE $=$ \gammaO).
\textbf{(d)}~Extraction of $\sqrt{|\Gamma_{\mathrm{o}}^{2}-\Gamma_{\mathrm{e}}^2|}$ from CSDs measured at various magnetic field angles.
\textbf{(e)} $G_{\mathrm{l}}$ measured at the centre of the CSD in (b), showing correlated zero-bias peaks. 
\textbf{(f)} $G_{\mathrm{l}}$ and $G_{\mathrm{nl}}$ measured upon detuning $V_{\mathrm{QDL}}$, while keeping $V_{\mathrm{QDR}}$ on resonance. 
Correlated zero-bias peaks persist across a large voltage range. 
Raw data and the extraction procedure is presented in Fig.~S3.
Measurements are taken at $B = $80 mT.
}
\label{fig2}
\end{figure}

\begin{figure*}[ht!]
\centering
\includegraphics[width=0.95\textwidth]{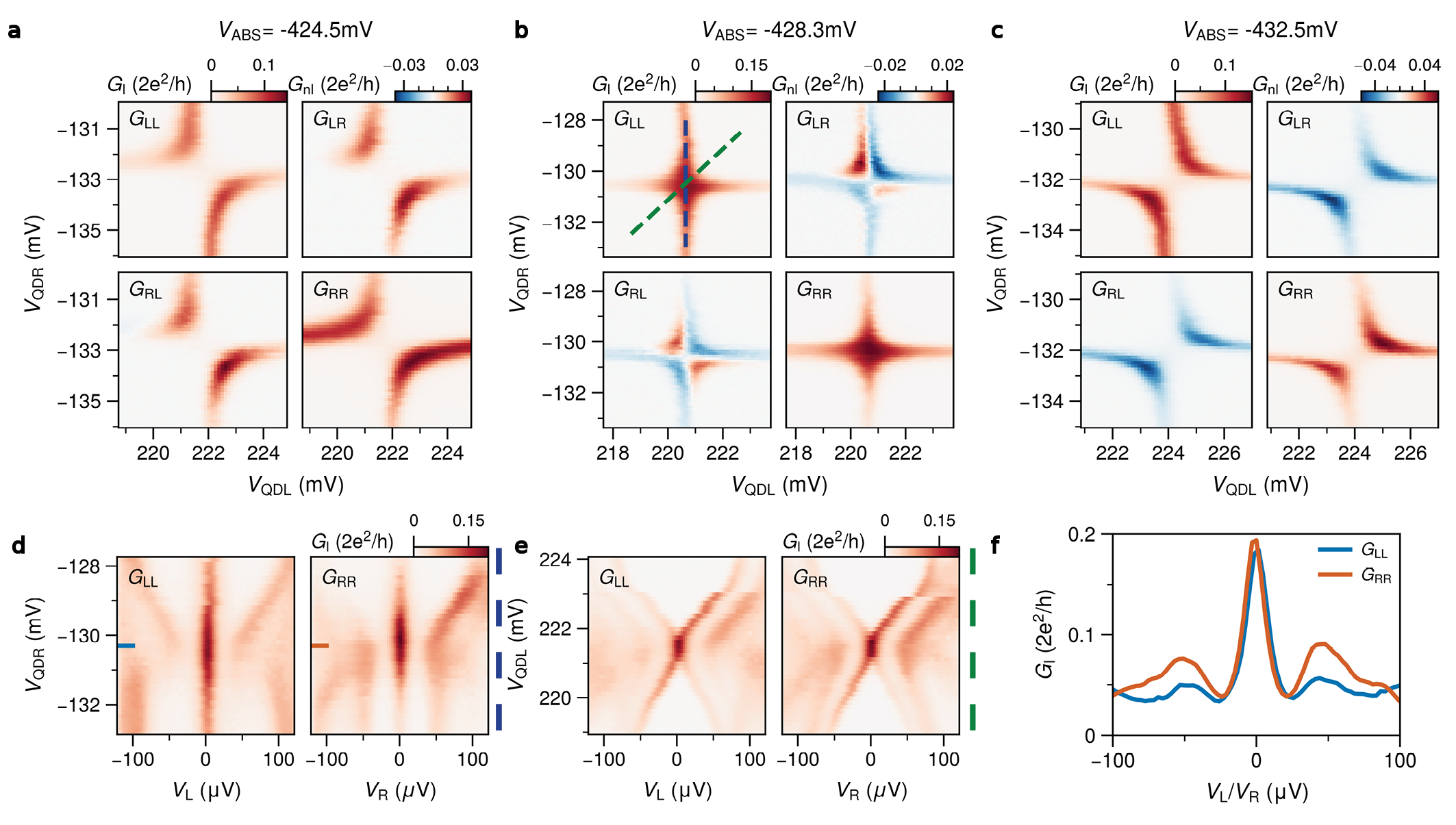}
\caption{\textbf{Electrostatically tuning to the Majorana sweet spot.} 
\textbf{(a)}, \textbf{(b)} and \textbf{(c)} show CSDs taken at three different applied voltages $V_{\mathrm{ABS}}$, in device A.
The system is smoothly tuned from the \gammaO $>$\gammaE~regime in (a) to the \gammaO $<$\gammaE~regime in (c). In between, the \gammaO $=$ \gammaE~condition is satisfied (b). A more extensive range is highlighted in Fig.~S6.
\textbf{(d)} Tunneling spectroscopy measurements at the sweet spot. $V_{\mathrm{QDR}}$ is tuned along the blue path shown in (b), while QDL is kept on resonance. 
\textbf{(e)} Tunneling spectroscopy as $V_{\mathrm{QDL}}$ and $V_{\mathrm{QDR}}$ are tuned simultaneously along the green path shown in (b). 
\textbf{(f)} Line-trace from (d) with $V_{\mathrm{QDL}}$ and $V_{\mathrm{QDR}}$ tuned to the sweet spot in (b) (corresponding to $\updelta\mu_{\mathrm{L}} = \updelta\mu_{\mathrm{R}} =0$). Data is taken with $B = \SI{150}{\mathrm{mT}}$. 
}
\label{fig3}
\end{figure*}

We demonstrate this control in \Cref{fig2}.
\Cref{fig2}a shows a CSD obtained with $B$~$\perp $~$B_{\mathrm{SO}}$, around the charge transition corresponding to the lower left corner of \Cref{fig1}d.
The diagonal avoided crossing indicates that \gammaE $>$\gammaO.
Rotating the field to align with $B_{\mathrm{SO}}$ results in an anti-diagonal avoided crossing, as now \gammaO $>$\gammaE~(\Cref{fig2}c). 
The separation between the branches of the avoided crossing is proportional to $\sqrt{|\Gamma_{\mathrm{o}}^2-\Gamma_{\mathrm{e}}^2|}$ (detailed in SI-Methods), which can be used to quantify the relative strength of the couplings. 
Measuring this quantity for several angles (\Cref{fig2}d) 
shows a smooth evolution of the coupling strength as a function of the field angle.
Importantly, at an intermediate angle the avoided crossing disappears~(\Cref{fig2}b), indicating \gammaE=\gammaO.
Under these conditions, the odd and even parity ground states are degenerate at the charge degeneracy point, i.e., $\updelta\mu_{\mathrm{L}} = \updelta\mu_{\mathrm{R}} = 0$, leading to localized MBSs on each QD~\cite{Liu2023}.
We refer to this point in parameter space as the Majorana sweet spot.
At the sweet spot, simultaneous tunneling spectroscopy on the left and right QD demonstrates correlated ZBPs (\Cref{fig2}e).
Higher energy excitations are visible at $\pm$~\SI{40}{\micro V}, providing an estimate of the effective couplings at the sweet spot to be \gammaE = \gammaO $\approx$ $\SI{20}{\micro eV}$.
These ZBPs are expected to persist when only a single QD is perturbed, as they result from MBSs localized on each of the QDs. 
To confirm this, we measure \loc~and \nloc~upon detuning $V_{\mathrm{QDL}}$, while keeping $V_{\mathrm{QDR}}$ constant (\Cref{fig2}f). 
The ZBPs indeed persist in \loc, while higher energy excitations are observed to disperse when QDL is detuned. 
Further, in \nloc~only the higher energy excitations are visible while the ZBPs themselves do not appear, a signature of the localized nature of these zero-energy states. 
These observations are consistent with experiments on nanowires~\cite{Dvir2023} and theoretical predictions~\cite{Leijnse2012}. \newline 
\begin{figure*}[t!]
\centering
\includegraphics[width=0.95\textwidth]{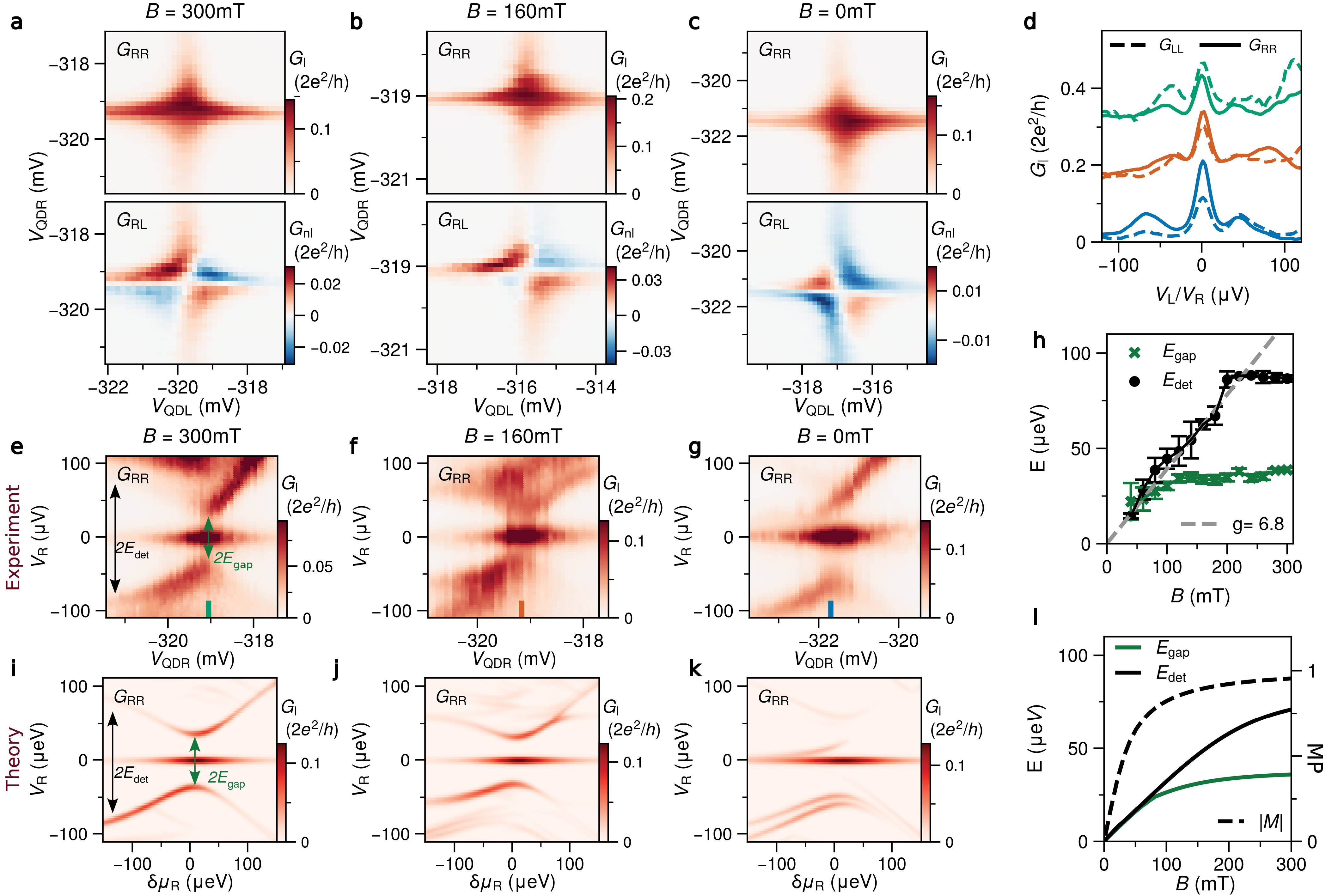}
\caption{\textbf{Majorana sweet spots in varying magnetic field.} 
\textbf{(a)}, \textbf{(b)} and \textbf{(c)} CSDs at ``Majorana sweet spots" measured at applied fields of \SI{300}{mT}, \SI{160}{mT} and \SI{0}{mT} respectively. 
At each field, $V_{\mathrm{ABS}}$ is adjusted to tune to the ``sweet spot", following the procedure from \Cref{fig3}.
\textbf{(d)} $G_{\mathrm{RR}}$ (solid) and $G_{\mathrm{LL}}$ (dashed) line-cuts at the centre of each CSD, from indicated positions in (e-g), highlighting the presence of correlated ZBPs. 
Offsets of 0.15 and 0.3 are applied for \SI{160}{mT} and \SI{300}{mT} respectively. 
\textbf{(e-g)}  Measured $G_{\mathrm{RR}}$ upon detuning QDR, while keeping QDL on resonance. Data is saturated for visibility of the excited states. Energies of interest ($E_{\mathrm{gap}}$ and $E_{\mathrm{det}}$) are highlighted in (e).
\textbf{(h)}  Extraction of the energies $E_{\mathrm{gap}}$ and $E_{\mathrm{det}}$ from measurements between \SI{300}{mT} and \SI{0}{mT} (full dataset in Fig.~S9). 
Dashed line shows a linear fit of $E_{\mathrm{det}}$, providing an estimate for the g-factor of QDL.
\textbf{(i-k)} Numerically calculated conductance, for qualitative comparison with (e-g). 
\textbf{(l)} Numerically extracted evolution of $E_{\mathrm{gap}}$ and $E_{\mathrm{det}}$ (solid) and corresponding Majorana polarization $|M|$ (dashed) as a function of the magnetic field. 
}
\label{fig4}
\end{figure*}

The above procedure for tuning to the sweet spot is guaranteed to work if one starts with a field angle where \gammaE $>$\gammaO, since \gammaE~can always be decreased by rotating the field towards $B_{\mathrm{SO}}$.
If a field rotation reveals that such an angle cannot be found, one can reach the sweet spot via electrostatic control over the hybrid section~\cite{Dvir2023}, since \gammaE~and \gammaO~are affected by the charge and energy of the ABSs~\cite{Liu2022, Bordin2022}.
Here, the ABS energies are controlled via the voltage $V_{\mathrm{ABS}}$, applied to the gate above the proximitized region.  With the magnetic field directed away from the spin-orbit field (as in \Cref{fig1}f), we study the evolution of the CSDs with $V_{\mathrm{ABS}}$. 
\Cref{fig3}a shows a diagonal avoided crossing, signifying \gammaO $>$\gammaE.
By tuning $V_{\mathrm{ABS}}$, the avoided crossing changes direction,  indicating \gammaE $>$ \gammaO~(\Cref{fig3}c).
At an intermediate $V_{\mathrm{ABS}}$ the avoided crossing disappears (\Cref{fig3}b), satisfying the sweet spot condition (\gammaE = \gammaO).
Similar to \Cref{fig2}, we now detune $V_{\mathrm{QDR}}$ along the blue dashed line in \Cref{fig3}b and measure $G_{\mathrm{l}}$, again finding correlated, persisting ZBPs (\Cref{fig3}d). 
Line-cuts from \Cref{fig3}d are shown in \Cref{fig3}, giving an estimate of \gammaE = \gammaO $=\SI{25}{\micro eV}$.
When both QDs are detuned simultaneously, along the green dashed line in \Cref{fig3}b, the correlated ZBPs disperse quadratically (\Cref{fig3}e). 
This is expected for a two-site Kitaev chain, where the ZBPs are only protected from local perturbations~\cite{Leijnse2012}.
An extended dataset and a comparison with numerical results is shown in Fig.~S5.

In addition to \loc, the non-local measurements in \Cref{fig3}a-c can provide further information about underlying transport mechanisms. 
For example, it has been shown that for CAR, local and non-local signals have the same sign, while for ECT their sign should be opposite \cite{Wang2022a}.
Since charge is ill-defined for YSR-states, there is no one-to-one correspondence between the dominant inter-dot coupling (i.e., \gammaE,~\gammaO) and the dominant underlying transport mechanism (i.e., CAR, ECT). 
Nevertheless we find a qualitatively similar behavior, whereby for~\gammaO$>$\gammaE~the non-local conductance is positive, whereas for \gammaE $>$\gammaO~it is negative.
We show that this is indeed expected (see Fig.~S7), and that the sign of $G_{\mathrm{L}}$ is dictated by the positions of the QDs w.r.t. their charge degeneracy.

\subsection{Majorana polarization}
The ideal Kitaev chain is based on a spinless model. 
Thus, emulating this system with spinful QDs, as presented here, requires the Zeeman energies of the QDs to be sufficiently large compared to the effective coupling between the QDs \cite{Leijnse2012, DasSarma2012, Fulga2013,tsintzis2023, Liu2023}.
In addition, MBSs on either QD should be isolated from each other.
A parameter capturing these factors is the so-called Majorana polarization (MP)~\cite{Bena2015}, which has recently been investigated theoretically in the context of Kitaev chains~\cite{Tsintzis2022}. 
The MP is a metric (denoted $|M|$) that quantifies the extent to which localized MBSs hybridize, and is relevant for experiments that require controlled manipulation of multiple MBSs, such as braiding~\cite{Boross2023, tsintzis2023} and parity-based qubits~\cite{Leijnse2012,Dai2015,Gabor2020,pino2023}.  
The MP ranges from 0 (lowest polarization) to 1 (highest polarization), and in experiments it is desirable to have a high value of MP.
It was shown that both ``low" and ``high" MP can result in similar transport signatures~\cite{Tsintzis2022}, raising an important question about how one could experimentally distinguish between these regimes. 

In order to investigate this, we track the evolution of the system from \SI{300}{mT} to \SI{0}{mT} (along $B_{\mathrm{z}}$).
At each field, we find similar crossings in the CSDs (\Cref{fig4}a-c). 
Simultaneous tunneling spectroscopy of the QDs at these crossing points reveal correlated ZBPs, down to zero magnetic field (\Cref{fig4}d). 
Furthermore, the behavior of non-local conductance around the centre of each crossing~\cite{Tsintzis2022} also shows no discernible differences as the field is reduced. 
At first instance, these observations are surprising, since MBSs require time reversal symmetry to be broken. 
However, we note that the combination of time-reversal symmetry and Coulomb interactions can result in robust zero energy modes associated with Kramers pairs of Majorana zero modes, as discussed in [32]. 
While further investigation is needed to confirm this interpretation, we can conclude that experimentally it is difficult to extract information about the MP from such measurements.

On the other hand, we find that the dispersion of higher energy excitations in tunnelling spectroscopy has a distinctly different behavior at each magnetic field, and allows us to obtain information about the MP.
While the ZBPs themselves persist upon detuning QDR for all values of $B$, the excited states show a markedly different behavior (\Cref{fig4}e--g). 
For example, at large detuning of QDR, excited states are visible at $\mathrm{\pm\SI{100}{\upmu V}}$ in \Cref{fig4}e, while reaching only $\mathrm{\pm\SI{60}{\upmu V}}$ in \Cref{fig4}f.
We denote this energy difference between the first excited states and the ZBPs as $E_{\mathrm{gap}}$ at the sweet spot and as $E_{\mathrm{det}}$ at large negative detuning, and extract these energies over an extensive range in $B$ (\Cref{fig4}h).
Both $E_{\mathrm{det}}$ and $E_{\mathrm{gap}}$ are found to increase monotonically with increasing magnetic field. 
For these sets of measurements, $E_{\mathrm{gap}}$ starts to saturate at $\mathrm{\SI{30}{\upmu V}}$ at higher fields, while $E_{\mathrm{det}}$ increases linearly until saturating at $\mathrm{\SI{80}{\upmu V}}$. 
The latter can be expected, since the excitation energy of the hybrid system approaches the excitation energy, i.e.,~the Zeeman energy of an isolated QD, when either of the QDs is detuned. 

We compare these measurements with numerical simulations of the system with input from experimental parameters, and find that indeed $E_{\mathrm{det}}$ provides a lower bound estimate of $E_{\mathrm{z}}$ of the QDs (Fig.~S8).
This allows to approximate a \textit{g}-factor from \Cref{fig4}h. 
Using this, the simulated spectra demonstrate a qualitatively similar behavior in the dispersion of the excited states (\Cref{fig4}i-k). 
Furthermore, we find that the evolution of both $E_{\mathrm{gap}}$ and $E_{\mathrm{det}}$ compare well to the experimental results~(\Cref{fig4}l). 

These results allow us to numerically estimate the Majorana polarization for the system, as a function of magnetic field (\Cref{fig4}l). 
We find that as $B$ increases, the MP increases quickly from 0 and then starts to approach 1 around \SI{100}{mT}, where $E_{\mathrm{gap}}$ begins to saturate. 
The comparison here yields $M\approx$ 0.96 at around $B = \SI{300}{mT}$. 
A similar analysis is performed for measurements using the sweet spot shown in \Cref{fig3}, where $E_{\mathrm{gap}}$ reaches $\mathrm{\SI{50}{\upmu V}}$, from which we extract a lower MP estimate of $M\approx$ 0.9~(Fig.~S10).
Whether these estimates can be considered a ``high" MP is dependent on the operations one intends to perform.
For example, specific braiding protocols have been shown to be reliable for $M = 0.98$ ~\cite{tsintzis2023}. It should be noted that these experiments do not constitute a direct measurement of the MP. One way to achieve this would be to introduce an additional QD on either side with a tunneling coupling to either MBS~\cite{Clarke2017,Prada2017,Souto2023}.
Regardless, the presented measurements show that the evolution of the system from zero magnetic field to high magnetic field can be well understood within the framework of the Kitaev model.   

\section*{Conclusion}
In summary, we have implemented a two-site Kitaev chain in a two-dimensional electron gas by coupling QDs through ABSs in a superconductor-semiconductor hybrid region. 
We demonstrate a smooth control over the inter-dot couplings, both by rotations of the magnetic field and by tuning the energy of Andreev bound states in the hybrid section. 
At specific points in the parameter space, zero energy excitations arise that are stable against local perturbations of either QD. 
We show that these ``sweet spots" (accompanied by correlated ZBPs) appear in the system even at zero Zeeman energy, and are by themselves insufficient to gain information about the polarization of MBSs. 
Rather, we find that the magnetic field dependence of the energy spectrum allows us to distinguish between high and low polarization regimes. 
Our work demonstrates that artificial Kitaev chains can now be realized on a scalable and flexible platform, and that 2DEGs are poised to perform Majorana-based experiments that were previously inaccessible.
\
\section{Acknowledgments}
We thank M. Leijnse, G. Katsaros, G. Wang, N. van Loo, A. Bordin and F. Zatelli for valuable discussions and for providing comments on the manuscript. The experimental research at Delft was supported by the Dutch National Science Foundation (NWO) and a TKI grant of the Dutch Topsectoren Program. 
A.M.B. acknowledges NWO (HOTNANO) for their research funding.
C-X.L. acknowledges subsidy by the Top consortium for Knowledge and Innovation program (TKI). S.G. and M.W acknowledge financial support from the Horizon Europe Framework Program of the European Commission through the European Innovation Council Pathfinder grant no. 101115315 (QuKiT).

\section{Author contributions}
S.L.D.t.H. fabricated the devices. Q.W. and I.K. contributed to the device design and optimization of fabrication flow. Measurements were performed by S.L.D.t.H., Q.W. and P.K. 
Numerical analysis was provided by S.L.D.t.H., A.M.B. and C-X.L, under the supervision of M.W.
MBE growth of the semiconductor heterostructures and the characterization of the materials was performed by D.X. and C.T. under the supervision of M.J.M. 
The manuscript was written by S.L.D.t.H., Q.W. and S.G., with inputs from all coauthors. T.D. and S.G. supervised the experimental work in Delft.

\section{Data availability}
Raw data and analysis scripts for all presented figures are available at \newline
\url{https://zenodo.org/records/10801215}. 

\bibliography{ref} 


\end{document}


\title{A two-site Kitaev chain in a two-dimensional electron gas}

\author{Sebastiaan L.D. ten Haaf}
\altaffiliation{These authors contributed equally to this work.}
\affiliation{QuTech and Kavli Institute of Nanoscience, Delft University of Technology, Delft, 2600 GA, The Netherlands}

\author{Qingzhen Wang}
\altaffiliation{These authors contributed equally to this work.}
\affiliation{QuTech and Kavli Institute of Nanoscience, Delft University of Technology, Delft, 2600 GA, The Netherlands}

\author{A. Mert Bozkurt}
\affiliation{QuTech and Kavli Institute of Nanoscience, Delft University of Technology, Delft, 2600 GA, The Netherlands}

\author{Chun-Xiao Liu}
\affiliation{QuTech and Kavli Institute of Nanoscience, Delft University of Technology, Delft, 2600 GA, The Netherlands}

\author{Ivan Kulesh}
\affiliation{QuTech and Kavli Institute of Nanoscience, Delft University of Technology, Delft, 2600 GA, The Netherlands}

\author{Philip Kim}
\affiliation{QuTech and Kavli Institute of Nanoscience, Delft University of Technology, Delft, 2600 GA, The Netherlands}

\author{Di~Xiao}
\affiliation{Department of Physics and Astronomy, Purdue University, West Lafayette, 47907, Indiana, USA}

\author{Candice Thomas}
\affiliation{Department of Physics and Astronomy, Purdue University, West Lafayette, 47907, Indiana, USA}

\author{Michael J. Manfra}
\affiliation{Department of Physics and Astronomy, Purdue University, West Lafayette, 47907, Indiana, USA}
\affiliation{Elmore School of Electrical and Computer Engineering, ~Purdue University, West Lafayette, 47907, Indiana, USA}
\affiliation{School of Materials Engineering, Purdue University, West Lafayette, 47907, Indiana, USA}

\author{Tom Dvir}
\affiliation{QuTech and Kavli Institute of Nanoscience, Delft University of Technology, Delft, 2600 GA, The Netherlands}

\author{Michael Wimmer}
\affiliation{QuTech and Kavli Institute of Nanoscience, Delft University of Technology, Delft, 2600 GA, The Netherlands}

\author{Srijit Goswami}\email{s.goswami@tudelft.nl}
\affiliation{QuTech and Kavli Institute of Nanoscience, Delft University of Technology, Delft, 2600 GA, The Netherlands}

\maketitle

\begin{widetext}
\tableofcontents
\clearpage
\section{Methods}

\subsection*{Device fabrication and yield}

All devices were fabricated using techniques described in detail in~\cite{Moehle2022}.
A narrow aluminum strip is defined in an InSbAs-Al chip by wet etching, followed by the deposition of two normal Ti/Pd contacts. 
After deposition of \SI{20}{nm} AlOx via atomic layer deposition (ALD), two Ti/Pd depletion gates are evaporated. 
Following a second ALD layer (\SI{20}{nm} AlOx), seven Ti/Pd finger gates are evaporated in order to define the QDs and tune the ABSs energy. 
The two depletion gates define a quasi-1D channel with a width of about \SI{160}{nm}, contacted on each side by a normal lead. The aluminium strip induces superconductivity in the middle section of each device, with an induced gap on the order of $\SI{200}{\upmu eV}$.
The presence of extended ABSs is confirmed through tunnelling spectroscopy. 
ABSs are found to be present over a large range of $V_{\mathrm{ABS}}$, the voltage applied to the gate covering the hybrid region.
Finger gates on the left and right of the aluminium define QDs with charging energies above \SI{1}{mV} (\Cref{figS1}a,b).

Two devices were used to obtain the data presented in the main text (device A and device B). Both showed strong hybridization between the QDs, as presented here. Device A was used to study the field evolution of Majorana sweet spots and to obtain the measurements presented in Fig.~1, Fig.~3 and Fig.~4. 
Device B was used to demonstrate the role of spin-orbit coupling on the interdot interactions, and the control over interactions through magnetic field as presented in Fig. 2.
Regarding device yield, up till now we have measured 12 devices for the purpose of studying hybridized QDs. Of these, we could tune 8 devices to regimes with strongly coupled QDs that showed the tunability displayed in Fig. 3. 
Of the non-functional devices, 2 failed due to trivial reasons (e.g. losing electronic connection due to missing bondwires after cooling).
The remaining 2 devices failed at the stage of forming the 1-D channel, where we found some optimization was needed to discover the optimal separation between the top and bottom depletion gate. Measurements were performed in a dilution refrigerator with a base temperature of \SI{20}{mK}. 

 \subsection*{Transport measurements and data processing}
Transport measurements are performed in AC and DC using a three-terminal set-up, where the aluminum is electrically grounded.
Each Ohmic lead is connected to a current-to-voltage converter and biased through a digital-to-analogue converter that applies both DC and AC biases.
Offsets of the applied voltage-bias on each side are corrected via independently calibrating the Coulomb peaks in the QDs on each side. 
The voltage outputs of the current meters are recorded with two digital multimeters and two lock-in amplifiers.  When applying a DC voltage to the left Ohmic ($V_{\mathrm{L}}$)  the right lead ($V_{\mathrm{R}}$) is kept grounded and vice versa. 
AC excitations are applied on each side with amplitudes around $\SI{5}{\upmu V}$ RMS and frequencies of \SI{19}{Hz} (left) and \SI{29}{Hz}  (right).
In this way, we measure the full conductance matrix $G$ by first measuring the response of $I_{\mathrm{L}}$ and $I_{\mathrm{R}}$ to $V_{\mathrm{L}}$ and then to $V_{R}$.
We account for the voltage-divider effect by correcting the conductances using known fridge line resistances $( \SI{3.6}{k\Omega}$ in device A,  $\SI{3.3}{k\Omega}$ in device B), as detailed in \cite{martinez2021}. 
This correction was done for all presented spectroscopy data, except for the data shown in Fig~3d-f. 
Magnetic fields were applied using a 3D vector magnet. 
The alignment of the magnetic field of device A  is expected to be accurate within ±10° and calibrated through performing tunneling spectroscopy of the hybrid section as a function of field angle. 
The alignment of device B is expected to be accurate within ±5°. 

Due to device instabilities or charge jumps, electrostatics of the QDs experience small drifts over the course of the measurements. 
Investigated orbitals were tracked while collecting the presented datasets. 
For each tunnelling spectroscopy measurements at a sweet spot, where $V_{\mathrm{QDL
}}$ and/or $V_{\mathrm{QDR}}$ were detuned, a CSD was obtained directly before and directly after to ensure that no drifts occurred during such a measurement. 
If such a drift occurred, the measurement was discarded and repeated. 
Such drifts are the cause of small discrepancies in gate voltages between highlighted paths in Fig.~3b and the measurements shown in Fig.~3d,e.
The highlighted paths represents the corrected path taken w.r.t. the CSD shown in Fig.~3b, based on the CSDs obtained before and after the measurements.

\subsection*{Extracting QD parameters}
In order to compare energy scales between experiments and numerical calculations, the gate voltages $V_{\mathrm{QDL}}$ and $V_{\mathrm{QDR}}$ are converted to electro-chemical potential energies $\mu_{\mathrm{L}}$ and $\mu_{\mathrm{R}}$.
For this purpose we extract the dimensionless lever arms $\alpha$. 
When forming sub-gap YSR states in the QDs, the effective lever arm of each QD around a zero-bias charge degeneracy can differ from the lever arm of the uncoupled QD, depending e.g. on hybridization with the hybrid region (Fig.~S4), extensively addressed in \cite{Zatelli2023}.
For analysis we therefore estimate both the normal-state lever arm (denoted $\alpha_{\mathrm{N}}$) and the lever arm of the sub-gap YSR-states (denoted $\alpha_{\mathrm{YSR}}$), at the specific $V_{\mathrm{ABS}}$ regimes of interest. 
Device B was operated in a regime without significant difference between $\alpha_{\mathrm{N}}$ and $\alpha_{\mathrm{YSR}}$, such that the analysis in Fig.~2d uses the lever arms extracted in Fig.~S2.
With the orbitals in Fig.~4, sweet spots were investigated at magnetic field values between \SI{0}{mT} and \SI{300}{mT}, where the energy of excited states $E_{\mathrm{det}}$ was extracted at a fixed detuning of $V_{\mathrm{QDR}}$. 
A g-factor is extracted from this data, by a linear fit of $E_{\mathrm{det}}$ up to 180mT (before saturation). 
For this dataset $\alpha_{\mathrm{YSR}}$ was only obtained at \SI{0}{mT} (Fig.~S1), such that any change in $\alpha_{\mathrm{YSR}}$ as a function of magnetic field can not be accounted for in the analysis in Fig.~4l and Fig.~4. 
When extracting $E_{\mathrm{det}}$ at constant detuning of $V_{\mathrm{QDR}}$ at each field, $\mu_{\mathrm{R}}$ may not be constant but rather is expected to decrease (Fig.~S4). 
We note that the lack of this correction may lead to slightly underestimating the g-factor in Fig.~4h, which in turn will lead to a lowered estimation of the Majorana polarization. 
For the extractions of $E_{\mathrm{det}}$ and $E_{\mathrm{gap}}$, $G_{\mathrm{LL}}$ and $G_{\mathrm{RR}}$ line-traces are obtained at the sweet spot and at a detuning of $V_{\mathrm{QDR}}$ of $-2\mathrm{mV}$, corresponding to a detuning of $\mu_{\mathrm{R}}\approx100\upmu \mathrm{e}V$. 
From each line-trace, the separation between the ZBPs and the first higher energy excitation is extracted by fitting Gaussian peaks symmetrically around zero-bias. 
Error-bars are given by the uncertainty in these fits.

\subsection{Numerical transport calculations} 
For all presented numerical results, a description of the system incorporating both the two QDs and the middle hybrid section was used, recently introduced in \cite{Tsintzis2022}.
We employ the same model and highlight key points and used numerical parameter below. 
The model considers tunneling between two normal QDs (L,R) and a central QD (M), which is in proximity to a superconductor.
Each site $j$ has electrochemical potential energy $\mu_{j}$.
A spin-conserving hopping $t$ allows transport between the outer QDs and central QD. 
The effect of spin-orbit interactions is included through a spin-flip hopping term $t_{\mathrm{SO}}$ between the outer QDs and central QD. 
The presence of the SC is included by attributing a superconducting pairing term $\Delta_{i}$ in each QD. 
To match the experimental geometry, $\Delta_{\mathrm{L,R}}, < \Delta_{\mathrm{M}}$
Lastly, the left and right sites are assigned a large on-site charging energy $U$ and a Zeeman splitting $E_{z}$ between the $\lvert\downarrow\rangle$ and $\vert\uparrow \rangle$ occupation. 
The Hamiltonian is constructed as follows:
\begin{align*}
    H=& \sum_{j,\sigma}{\mu_j n_{j\sigma}} + \sum_{j}U_{j}n_{j\uparrow }n_{j\downarrow} + \sum_{j}{E_Z}_j(n_{j\downarrow} -n_{j\uparrow})\\
    +&\sum_{j\neq \mathrm{M},\sigma}[t_{j}d_{j\sigma}^{\dagger}d_{C\sigma}+h.c.]\\
    +&\sum_{j\neq \mathrm{M}}[t_{j}^{\mathrm{SO}}d_{j\downarrow}^{\dagger}d_{C\uparrow}-t_{j}^{\mathrm{SO}}d_{j\uparrow}^{\dagger}d_{C\downarrow}+h.c.]\\
    +&\sum_{j}[\Delta_{\mathrm{j}} d_{j\uparrow}^{\dagger}d_{j\downarrow}^{\dagger}+h.c.]
\end{align*}
where $d_{j\sigma},d_{j\sigma}^{\dagger}$ and $n_{j\sigma}$ are the annihilation, creation, and number operators respectively for each site.
The sum $j$ runs over the sites (L,M,R) and $\sigma$ over the spin degree of freedom ($\uparrow,\downarrow$).
For simplicity, a left and right symmetry is assumed, such that a Majorana sweet spot lies along $\mu_{\mathrm{L}}=\mu_{\mathrm{R}}$. 
Sweet spots are obtained by scanning the parameter space spanned by $\mu_{\mathrm{L}}=\mu_{\mathrm{R}}$ and $\mu_{\mathrm{M}}$ and finding degeneracies between lowest odd and even eigenstates.
All transport calculations are obtained using the rate equation detailed in~\cite{Tsintzis2022}.
Similar to \cite{Tsintzis2022}, we calculate the quantity of Majorana polarization (MP) (originally introduced in \cite{Bena2015}).
This quantity, denoted $M_j$, is defined per site $j$ as:
\begin{align*}
    M_{j} &= \frac{\sum_{\sigma} (w_{\sigma}^2 - z_{\sigma}^2)}{\sum_{\sigma}(w_\sigma^2+z_\sigma^2)} \\
    w_\sigma &= \langle O\lvert (d_{j\sigma} +d_{j\sigma}^\dagger)\lvert E \rangle \\
    z_\sigma &= \langle O\lvert (d_{j\sigma} -d_{j\sigma}^\dagger)\lvert E \rangle
\end{align*}
where $\lvert O\rangle$ and $\lvert E\rangle$ are the lowest-energy odd and even states respectively.
Due to symmetrically chosen parameters $\lvert M_{\mathrm{L}}\rvert=\lvert M_{\mathrm{R}}\rvert = |M|$, such that a single MP value can be extracted for a specific set of parameters.
To provide an experimental observable that reflects a high or low $|M|$, we investigate the behavior of two transition energies, denoted in the main text as $E_{\mathrm{gap}}$ and $E_{\mathrm{det}}$.
Numerically these are obtained from the Hamiltonian as the energy difference between the lowest even state and the second-lowest odd state, at specific values of $\mu_{\mathrm{L}}$ and $\mu_{\mathrm{R}}$.
$E_{\mathrm{gap}}$ is obtained with $\mu_{\mathrm{L}}$ and $\mu_{\mathrm{R}}$ set to corresponding to their sweet spot value. $E_{\mathrm{det}}$ is obtained with $\mu_{R}$ detuned by $\approx\Delta_{\mathrm{M}}$ .
In the absence of the Zeeman term, both $E_{\mathrm{gap}}$ and $E_{\mathrm{det}}$ are zero by definition, due to the degeneracy of the odd states in the presence of time reversal symmetry.
The dependence of $M$, $E_{\mathrm{gap}}$ and $E_{\mathrm{det}}$ on $E_{\mathrm{z}}$ is demonstrated in Fig.~S10. 
 \\

In order to compare to experiments, parameters are selected to match realistic values.
We set $\Delta_{\mathrm{M}}$ = $\SI{100}{\mathrm{\upmu e}V}$ and $\Delta_{\mathrm{L,R}} = 0.5\Delta_{\mathrm{M}}$. 
The charging energy $U$ is fixed in both dots to be 10$\Delta_{\mathrm{M}}$, except for Fig.~1d where a value of 7$\Delta_{\mathrm{M}}$ is used to better highlight the behavior of all four avoided crossings in the large CSD. 
For the analysis in Fig.~4, tunneling terms $t$ and $t_{\mathrm{SO}}$ are fine-tuned such that at large $E_{\mathrm{z}}$ the sweet spot gap of $\SI{30}{\upmu eV}$ is obtained, to match the experimental result.  
This gives $t_{\mathrm{SO}} = 0.4t = 0.7\Delta_{\mathrm{M}}$.
Similarly, to compare to the experimental results in Fig.~S5 and Fig.~S10, $t$ and $t_{\mathrm{SO}}$ are fine-tuned to obtain a sweet spot gap of $\SI{50}{\upmu eV}$ at large $E_{\mathrm{z}}$. 
This results in $t_{\mathrm{SO}} = 0.4t = 0.85\Delta_{\mathrm{M}}$.
\newline


\subsection{Effective model in the strong coupling regime}
The above description is used for all presented calculations.
To provide an intuitive understanding of these results, we invoke a description of the system through an effective model, introduced in detail in \cite{Liu2023}.
Here we provide a brief summary of the relevant findings.
When the energy $E_{\mathrm{ABS}}$ of the sub-gap state in the middle site is large compared to tunnel couplings between the QDs, the middle site can be integrated out.
This leaves a description of the system including only effective couplings between YSR states in the left and right QD sites. 
Additionally, it is assumed that $E_{\mathrm{z}}$ is sufficiently large such that only the $\lvert\downarrow\rvert$ occupation of each QD partakes in transport. 
In this description, a Hamiltonian can be constructed in a singlet-doublet basis. 
The ground state of each QD is either the $\lvert \downarrow \rangle$ doublet or a singlet $\lvert S \rangle$ of the form $\lvert S\rangle= u_{\mathrm{L,R}}\lvert 0 \rangle -v_{\mathrm{L,R|}}\lvert \downarrow\uparrow\rangle$.
Here the $u,v$ components depend on the chemical potential energies $\mu_{\mathrm{L}}$ and $\mu_{\mathrm{R}}$ of the left and right dot: ($u_{\mathrm{L,R}}^2 = 1-v_{\mathrm{L,R}}^2 = \frac{1}{2}+\frac{\mu_{L,R} + U}{\sqrt{U^2+E_{\mathrm{ABS}}^2}}$). 
The effective Hamiltonian is obtained:
\begin{equation}
    H^{\text{eff}}_{\text{coupling}} =\sum_{\sigma, \eta=\uparrow, \downarrow} (t_{\sigma \eta} c^{\dagger}_{L\sigma} c_{R\eta} + \Delta_{\sigma \eta} c^{\dagger}_{L\sigma} c^{\dagger}_{R \eta}) + h.c.,
\end{equation}
where $c_{\mathrm{L,R}}$ and $c_{\mathrm{L,R}}^{\dagger}$ are annihilation and creation operators for the left and right sites. Further, $t_{\mathrm{\sigma\eta}}$ specifies the amplitude of an electron with spin $\sigma$ from left site tunneling to occupy a state with spin $\eta$ in the right side, or vice versa.
$\Delta_{\mathrm{\sigma\mu}}$ specifies the amplitude of the creating or breaking of a Cooper pair through electrons with spins $\sigma$ and $\eta$ in the left and right QDs ($\sigma,\eta\in[\uparrow,\downarrow]$). 
Due to the spin-orbit interaction two equal-spin electrons can also be coupled via the $\Delta$-term.
Importantly, this description is equivalent to the weakly coupled model discussed in~\cite{Leijnse2012}, when changing from a charge occupation basis to a YSR-state basis.
Now, two types of coupling arise between the total even-parity states (denoted $\Gamma_{\mathrm{E}}$) and between total-odd parity states ($\Gamma_{\mathrm{O}}$).
Each are combination of ECT and CAR amplitudes:

\begin{equation*}
\begin{aligned}
\Gamma_{\mathrm{E}} &= \langle S S\lvert H^{\text{eff}}_{\text{coupling}}  \lvert\downarrow \downarrow\rangle
&=  -\Delta_{\su \su} v_L v_R + \Delta_{\downarrow \downarrow} u_L u_R  + t_{\su \downarrow} v_L u_R -t_{\downarrow \su} u_L v_R,\\
\Gamma_{\mathrm{O}} &= \langle S\downarrow\rvert H^{\text{eff}}_{\text{coupling}}  \lvert\downarrow S\rangle
&= -t_{\su \su} v_L v_R + t_{\downarrow \downarrow} u_L u_R + \Delta_{\su \downarrow} v_L u_R -\Delta_{\downarrow \su} u_L v_R,
\end{aligned}
\end{equation*}

The coupling $\Gamma_{\mathrm{O}}$ results in bonding and anti-bonding states of the form $u\lvert \mathrm{S},\downarrow\rangle\pm v\lvert \downarrow,\mathrm{S}\rangle$.
$\Gamma_{\mathrm{E}}$ forms bonding and anti-bonding states of the form $\alpha \lvert\mathrm{S},\mathrm{S}\rangle \pm \beta \lvert \downarrow,\downarrow \rangle$.
When $\Gamma_{\mathrm{E}} = \Gamma_{\mathrm{O}}$ the even and odd ground states become degenerate at $\mu_{\mathrm{L}}=\mu_{\mathrm{R}}$, equivalent to the $t=\Delta$ condition in the "poor Man's Majorana" description~\cite{Leijnse2012}.
In Fig.~2d, we extract the distance between the branches of the avoided crossing in the experimentally obtained CSDs, after converting the gate voltages to the YSR energies of each QD.
In the effective description this distance corresponds to analysing where the line $\mu_{L}=\mu_{R}$ or $\mu_{L} = -\mu_{R}$ intersects a degeneracy between the even and odd ground state energies. 
This gives two points, separated by $\sqrt{8|\Gamma_{\mathrm{O}}^2-\Gamma_{\mathrm{E}}^2|}$.

\setcounter{figure}{0}
\renewcommand{\thefigure}{S\arabic{figure}}
\newpage
\section{Supplementary Figure S1 to S10}
\begin{figure}[h!]
\centering
\includegraphics[width=0.95\textwidth]{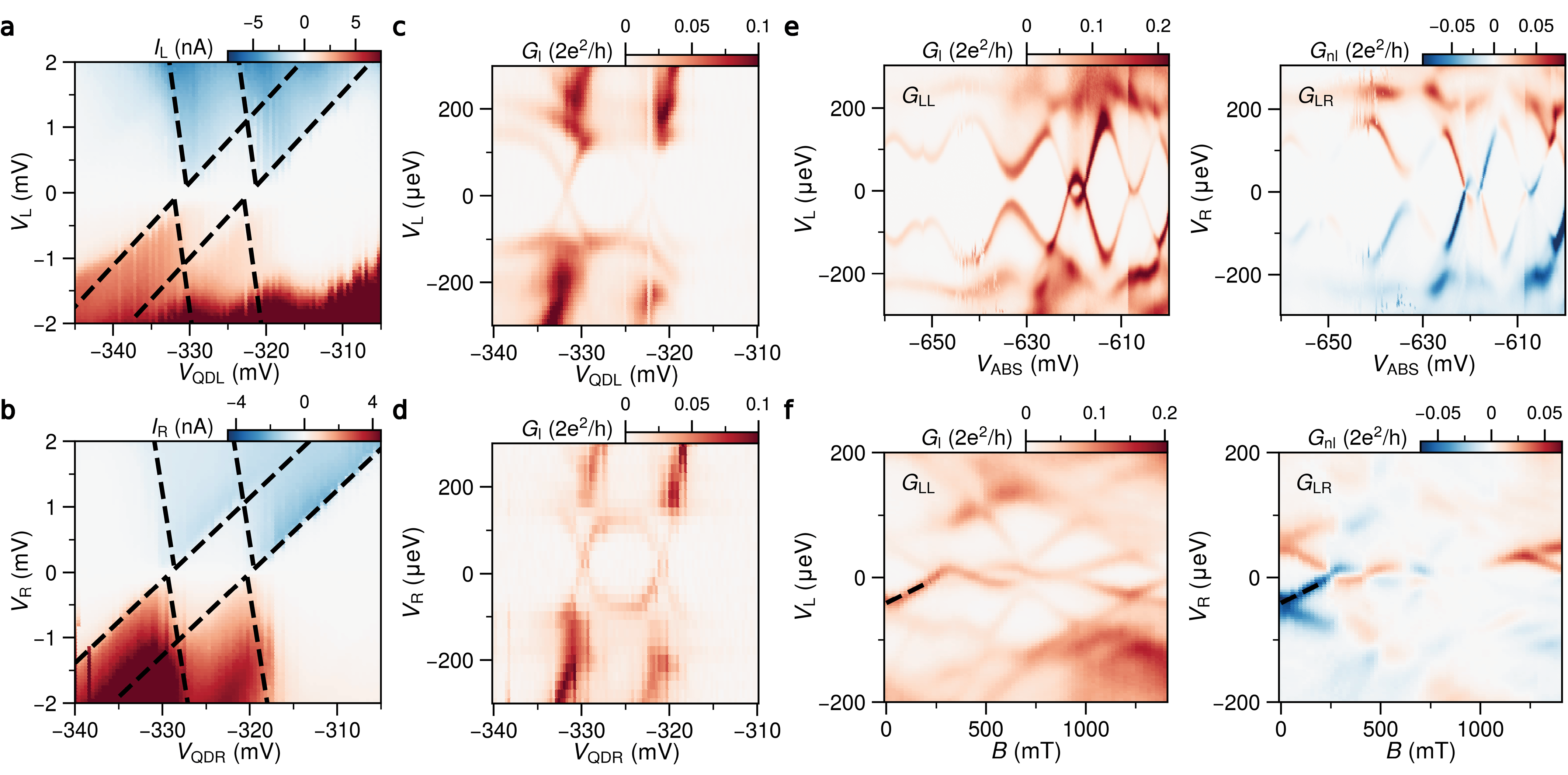}
\caption{\textbf{Characterization of QDs and ABS spectroscopy for device A.}
\textbf{(a-b)} Coulomb diamond measurements of the left and right QDs, in the regime used for measurements presented in Fig.~4.
Charging energies are estimated to be \SI{1.1}{meV} in each QD. 
The lever arm of the left and right dot is estimated to be $\alpha \approx$~0.11. 
\textbf{(c-d)} Zoomed in views of figures (a) and (b) in a smaller energy range. Due to a large tunnel coupling between the QDs and the hybrid section, YSR-states form in the sub-gap spectrum of the QDs. 
Tunneling spectroscopy around the charge degeneracy points in (a-b) reveal clear sup-gap features within the Coulomb diamonds. As $V_{\mathrm{QDL}}$ and $V_{\mathrm{QDR}}$  are tuned, the sub-gap features form an eye-shape feature enclosing the doublet charge occupation. This behavior is typical for YSR-states with large charging energies~\cite{Rasmussen2018}. 
\textbf{(e)} Crossed Andreev reflection and elastic co-tunneling require the presence of extended ABSs.
Local $G_{\mathrm{l}}$ and non-local conductance $G_{\mathrm{nl}}$ of the hybrid region are measured via tunnelling spectroscopy and their identical energy dependence as a function of $V_\mathrm{ABS}$  highlights that ABSs extend across the entire hybrid section. 
Comparable behavior was observed in a wide $V_{\mathrm{ABS}}$ range from 0 to \SI{-1}{V}.
The measurement presented in Fig.~1 is taken at the $V_{\mathrm{ABS}}$ with the eye-shaped crossing.
\textbf{(f)} ABS spectroscopy as a function external magnetic field at $V_{\mathrm{ABS}}$ = \SI{-623}{mV}. 
The effect of splitting of the doublet state can be observed at low fields.
A $g$-factor of 5.5 is extracted by linear fitting of the lowest sub-gap states (dashed line) in \Cref{figS1}d.
}
\label{figS1}
\end{figure}
\newpage

\begin{figure}[h!]
\centering
\includegraphics[width=0.95\textwidth]{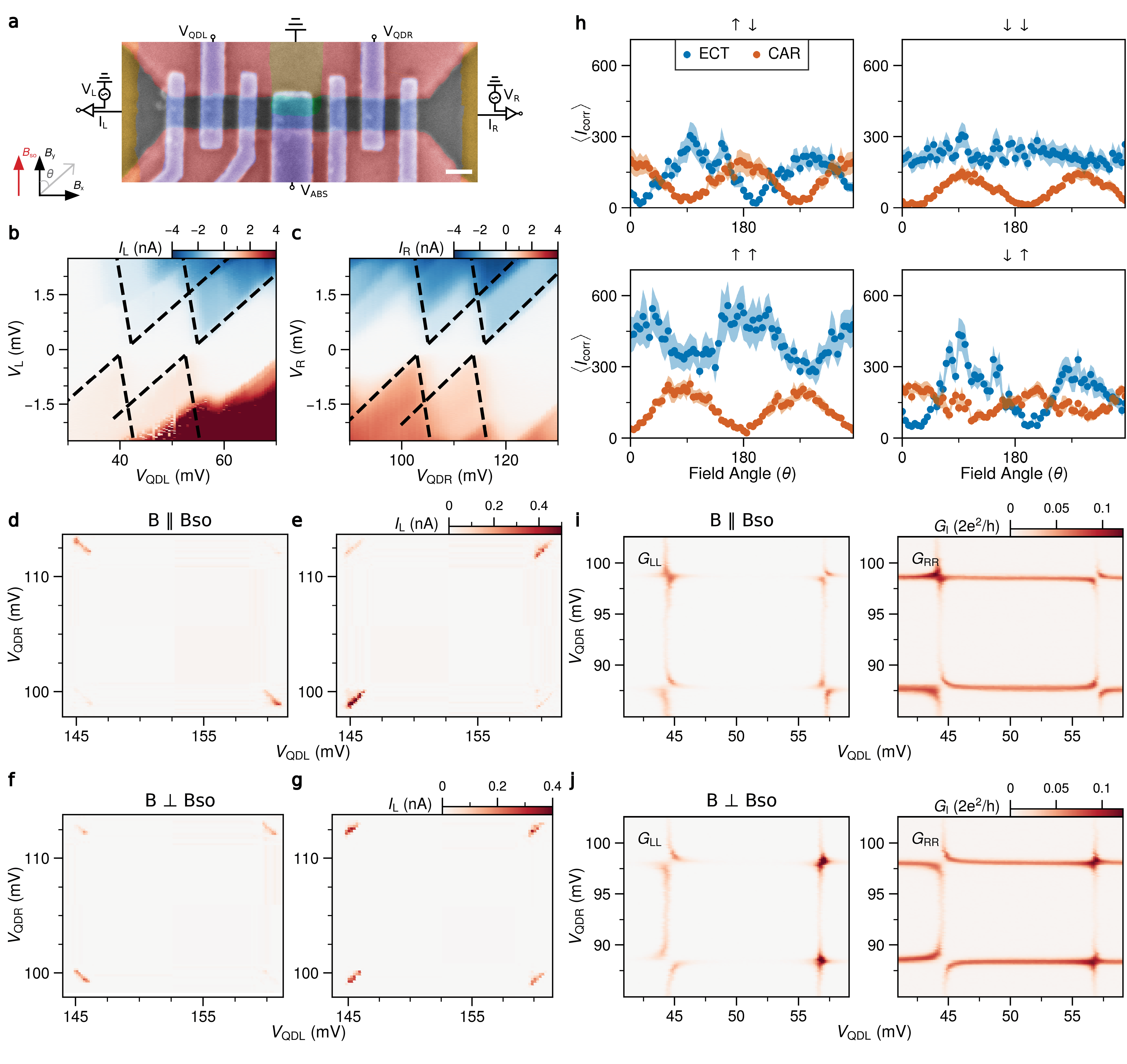}
\caption{\textbf{Characterization of device B.} \textbf{(a)} Scanning electron micro-graph of device B, used to obtain the measurements presented in Fig.~2. Scale bar shown is \SI{100}{nm}.
\textbf{(b-c)} Coulomb diamond measurements of the left and right QDs. Charging energies are extracted to be \SI{1.4}{meV}. The lever arm $\alpha_{\mathrm{N}}$ is extracted to be about 0.11 for each QD. This lever arm is used for the extraction in Fig.~2d.
(\textbf{d-e}) To validate the direction of $B_{\mathrm{SO}}$ and to show the connection between interactions in strongly coupled QDs and underlying ECT and CAR processes, we first measure ECT and CAR currents in the weakly coupled dots, as detailed in~\cite{WangQ2022}.
With $B\perp B_{\mathrm{SO}}$, measurements of CAR (d) and ECT (e) show the typical blockades for same-spin and opposite-spin charge configurations respectively. 
In (\textbf{f-g}), with $B\parallel B_{\mathrm{SO}}$, the spin non-conserving ECT and CAR processes are observed to be revived. 
\textbf{(h)} Measuring CAR and ECT rates as a function of magnetic field angle $\theta$ shows the currents for the spin non-conserving processes are indeed smoothly controlled and become suppressed when $\theta = 0$.
This supports the interpretation that $B\parallel B_{\mathrm{SO}}$ when B is perpendicular to the 1-D channel. 
\textbf{(i-j)} Next, the QDs are operated with higher tunnelling rates between the QDs and the SC, to enable strong couplings. 
Similar to Fig.~1e-f, CSDs are obtained in the strongly interacting regime, taken with the verified $B\parallel B_{\mathrm{SO}}$ and $B\perp B_{\mathrm{SO}}$ respectively. 
}
\label{figS2}
\end{figure}

\newpage

\begin{figure}[h!]
\centering
\includegraphics[width=0.95\textwidth]{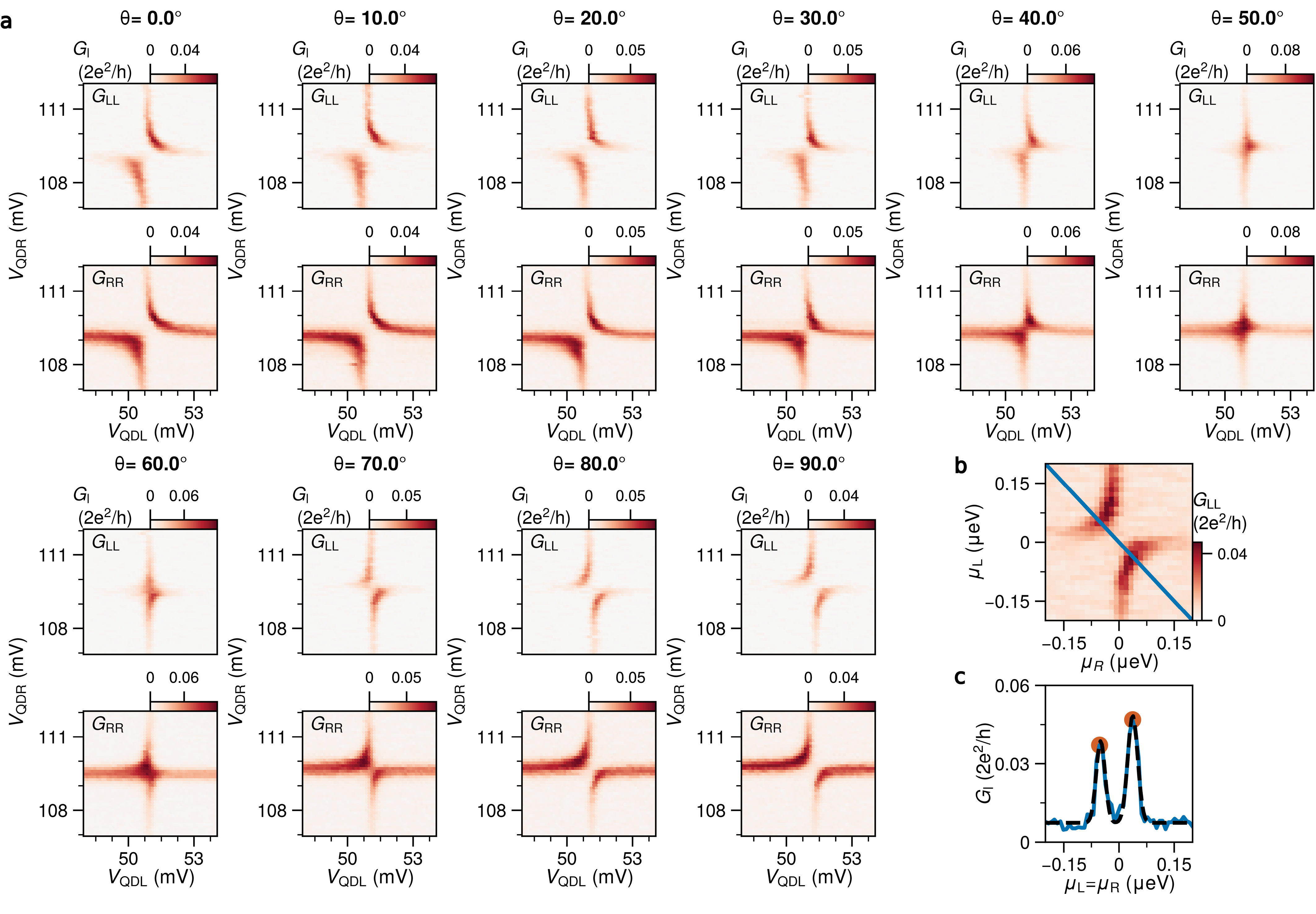}
\caption{\textbf{Extended dataset of Figure 2}.  \textbf{(a)} CSDs measured at various magnetic field angles $\theta$ between $\SI{0}{\degree}$ and $\SI{90}{\degree}$, used to extract the data shown in Fig.~2d. \textbf{(b)} Example of the extraction process. For each obtained CSD, $V_{\mathrm{QDL}}$ and $V_{\mathrm{QDR}}$ are converted to energies $\mu_{\mathrm{L}}$ and $\mu_{\mathrm{R}}$ using lever arms obtained in Fig.~S2. 
Next, the conductance is extracted along a $\mu_\mathrm{L} = -\mu_\mathrm{R}$ or $\mu_\mathrm{L}=\mu_{\mathrm{R}}$ line. \textbf{(c)} Two Gaussian peaks are fitted  to extract the separation between the two avoided crossings, from which the quantity $\sqrt{|\Gamma_{\mathrm{o}}-\Gamma_{\mathrm{e}}|}$ is obtained (plotted in Fig.~2d). 
}
\label{figS3}
\end{figure}

\begin{figure}[h!]
\centering
\includegraphics[width=0.95\textwidth]{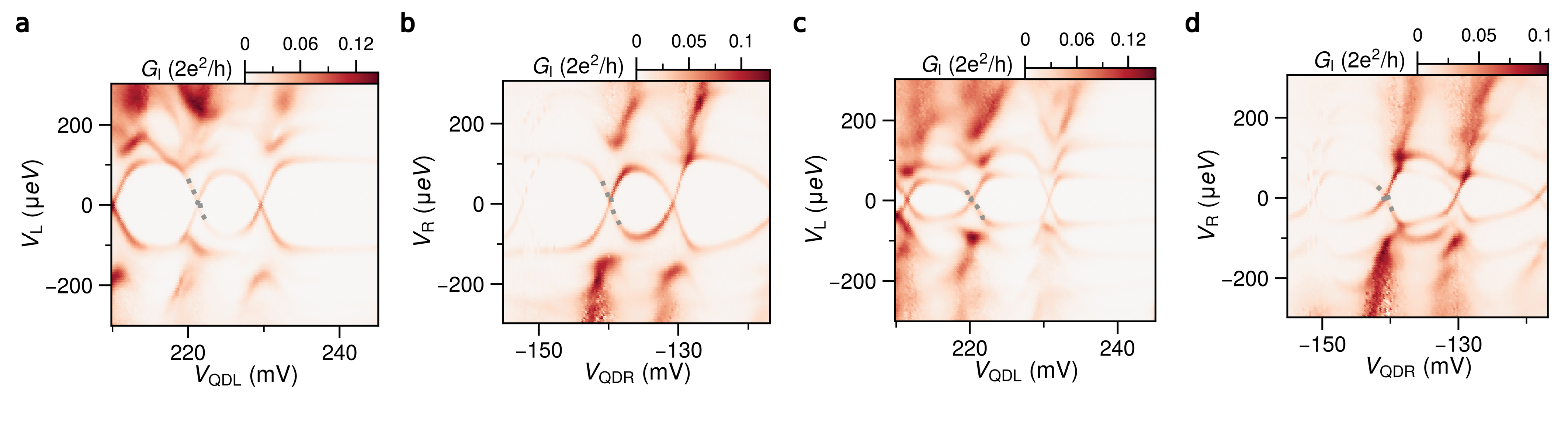}
\caption{\textbf{Characterisation of YSR-states in QDs of Figure 3.} 
To complement the data in Fig.~3, Fig~S5 and Fig.~S10, we measure the sub-gap states in QDL and QDR (see Fig~S1). 
Using this, we obtain the lever arms of $V_{\mathrm{QDL}}$ and $V_{\mathrm{QDR}}$ on the YSR-state energies (denoted $\alpha_{\mathrm{YSR}}$) (see Methods). 
\textbf{(a-b)} Sub-gap spectroscopy of QDL and QDR at $B_{\mathrm{z}}$=\SI{0}{mT}. 
From the slopes of the states upon crossing $V_{\mathrm{L}},V_{\mathrm{R}}=0$, we estimate $\alpha_{\mathrm{YSR}} \approx$ 0.045.
Applying an external magnetic field lowers the energy of ABSs in the hybrid region, as a result of Zeeman splitting,
This in turn will affect the YSR-spectrum of the QDs, due to increased hybridization between the QDs and the ABS. 
\textbf{(c-d)} Measuring sub-gap spectroscopy of QDL and QDR at $B_{\mathrm{z}}$=\SI{225}{mT} for the same settings as in (a-b) shows indeed the effective lever arm here decreases to $\alpha_{\mathrm{YSR}}\approx$ 0.028.}
\label{figS4}
\end{figure}

\begin{figure}[h!]
\centering
\includegraphics[width=0.95\textwidth]{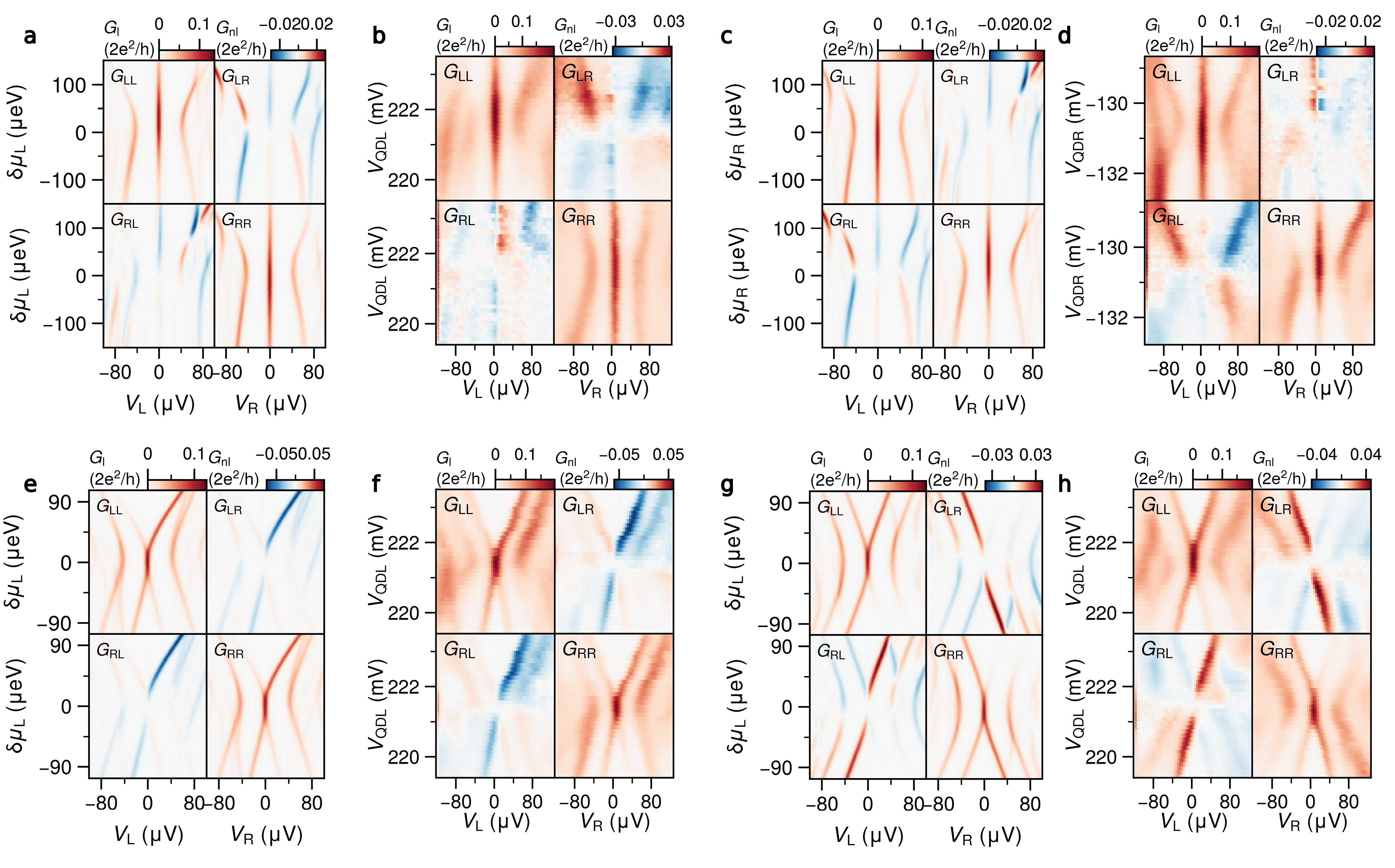}
\caption{\textbf{Full conductance spectra at the sweet spot upon detuning $V_{\mathrm{QDL}}$ and $V_{\mathrm{QDR}}$.} A comparison between numerically calculated conductance and measured conductance in support of Figure~3, measured at $B=\SI{225}{mT}$. Presented results show the evolution of $G_{\mathrm{l}}$ and $G_{\mathrm{nl}}$ for four different cases: \textbf{(a-b)} detuning $V_{\mathrm{QDL}}$,   \textbf{(c-d)} detuning $V_{\mathrm{QDR}}$,  \textbf{(e-f)} detuning both $V_{\mathrm{QDL}}$ and $V_{\mathrm{QDR}}$ simultaneously along a diagonal path and \textbf{(g-h)} detuning both anti-diagonally.
For each case, we find the behavior of both $G_{\mathrm{l}}$ and $G_{\mathrm{nl}}$ is well described by the numerical results. 
}
\label{figS5}
\end{figure}

\begin{figure}[h!]
\centering
\includegraphics[width=0.95\textwidth]{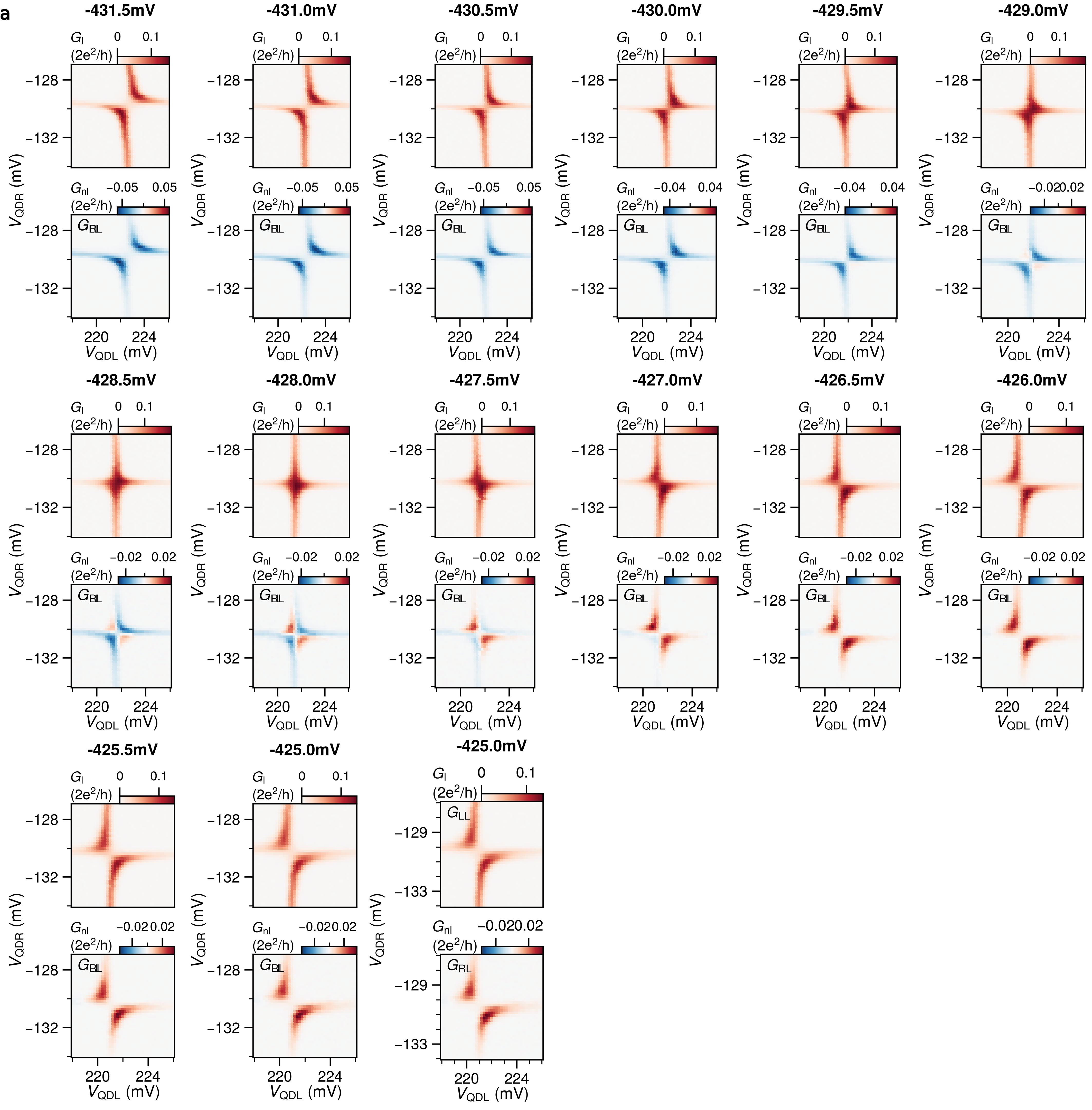}
\caption{\textbf{Extended dataset for Figure 3.} \textbf{(a)}  Sets of CSDs obtained while varying $V_{\mathrm{ABS}}$ in the range presented in Fig~3. The range of $V_{\mathrm{QDL}}$ and $V_{\mathrm{QDR}}$ is constant for each measurement. 
The slight drift of the avoided crossing upon varying $V_{\mathrm{ABS}}$ is owed to cross-capacitance between $V_{\mathrm{ABS}}$ and the potential of the QDs. 
}
\label{figS6}
\end{figure}

\begin{figure}[h!]
\centering
\includegraphics[width=\textwidth]{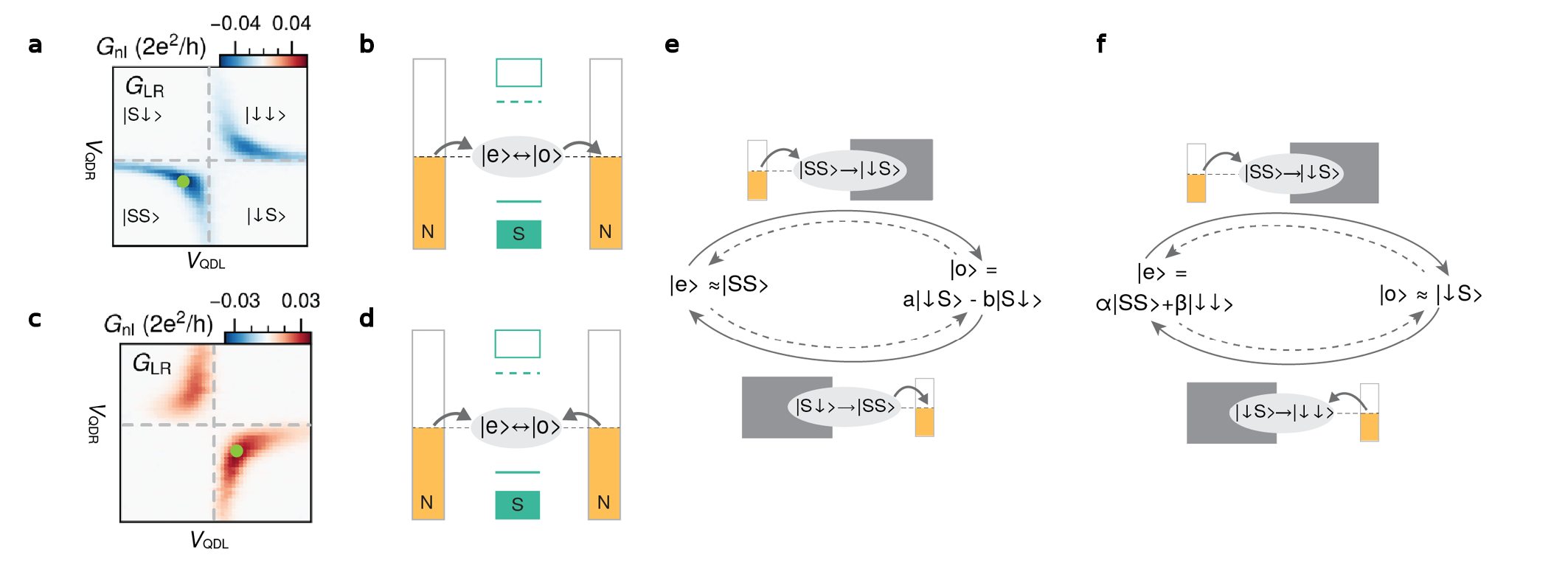}
\caption{\textbf{Energy diagrams detailing  non-local transport.} In CSDs presented in Fig.~3, a clear sign change is observed when changing from the $\Gamma_{\mathrm{O}}$$>\Gamma_{\mathrm{E}}$ regime to the $\Gamma_{\mathrm{O}}$$<\Gamma_{\mathrm{E}}$ regime. This can be understood by considering the possible transport cycles that underlie the measured non-local conductance. \textbf{(a)} When $\Gamma_{\mathrm{O}}$$>\Gamma_{\mathrm{E}}$, $G_\mathrm{nl}$ is observed to be negative in the measured CSDs (see Fig.~3c). \textbf{(c)}  When $\Gamma_{\mathrm{O}}$$<\Gamma_{\mathrm{E}}$, the same measurements yield a positive $G_\mathrm{nl}$ (see Fig.~3a). 
Horizontal and vertical dashed lines indicate $\upmu_\mathrm{R}$ = 0 and $\upmu_\mathrm{L}$ = 0 respectively. The state of the uncoupled system is labelled in each quadrant. 
\textbf{(b,d)} In such CSD measurements, zero-bias transport can take place when the odd and even ground states are degenerate. For non-local transport to occur, the system can accept a hole/electron from one lead, and relax non-locally to its original state by either (b) donating a hole/electron to the opposite lead, giving rise to negative $G_\mathrm{nl}$,  or (d) accept a hole/electron from the opposite lead, giving rise to positive $G_\mathrm{nl}$.
The preferred path is dictated by the quadrant in $\mu_{\mathrm{L}}$, $\mu_{\mathrm{R}}$ space where the odd-even degeneracy occurs. 
\textbf{(e)} When $\mu_{\mathrm{L}}$, $\mu_{\mathrm{R}}>0$ or $\mu_{\mathrm{L}}$,$\mu_{\mathrm{R}}<0$, the former path is expected to dominate and the resulting $G_\mathrm{nl}$ will be negative. 
\textbf{(f)} When $\mu_{\mathrm{L}}>0$ and $\mu_{\mathrm{R}}<0$ or vice versa, the latter path is expected to dominate and resulting $G_\mathrm{nl}$ will be positive. }
\label{figS7}
\end{figure}

\begin{figure}[h!]
\centering
\includegraphics[width=0.95\textwidth]{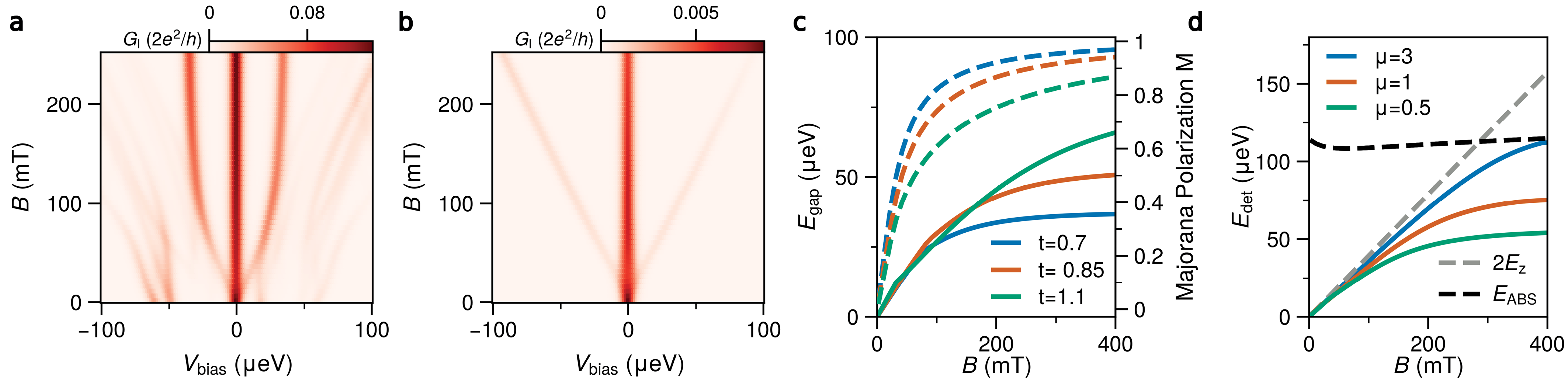}
\caption{\textbf{Numerical analysis of $E_{\mathrm{gap}}$ and $E_{\mathrm{det}}$} Numerical calculations supporting the results presented in Figure 4. 
Through the procedure detailed in SI-Methods, Majorana sweet spots are obtained and analysed for fields between \SI{0}{mT} and \SI{300}{mT}.
\textbf{(a)} Field evolution of $G_{\mathrm{RR}}$ line-traces at each sweet spot, showing the excitations above the ZBPs gradually increasing in energy and then saturating at $\SI{\pm 30}{\micro eV}$. 
\textbf{(b)} Field evolution of $G_{\mathrm{RR}}$ line-traces when QDR detuned by $3\Delta_{\mathrm{ind}}$, showing the excited states increase linearly in energy.
From calculations in (a) and (b), $E_{\mathrm{gap}}$ and $E_{\mathrm{det}}$ are obtained, given by the energy between the lowest even-parity state and second-lowest odd-parity state.
\textbf{(c)} Extraction of $E_{\mathrm{gap}}$ (solid) and the Majorana polarization (dashed), for different values of the tunneling parameter $t$. In each case $t_{so} = 0.4t$. Larger tunnel coupling results in larger hybridization between ABSs, in turn lowering the MP at a specific magnetic field. 
\textbf{(d)} Extraction of $E_{\mathrm{det}}$ for various values of detuning $\upmu_{R}$.
In each case the slope at low fields corresponds to $2E_{\mathrm{z}}$ (dashed grey line). The larger the detuning of $\upmu_{R}$, the longer this holds. The dashed black line shows the energy of the ABS $E_{\mathrm{ABS}}$. At large detuning $E_{\mathrm{det}}$ will increase linearly with $2E_{\mathrm{z}}$, until becoming of comparable $E_{\mathrm{ABS}}$ becomes the lowest energy scale. }
\label{figS8}
\end{figure}

\begin{figure}[h!]
\centering
\includegraphics[width=0.9\textwidth]{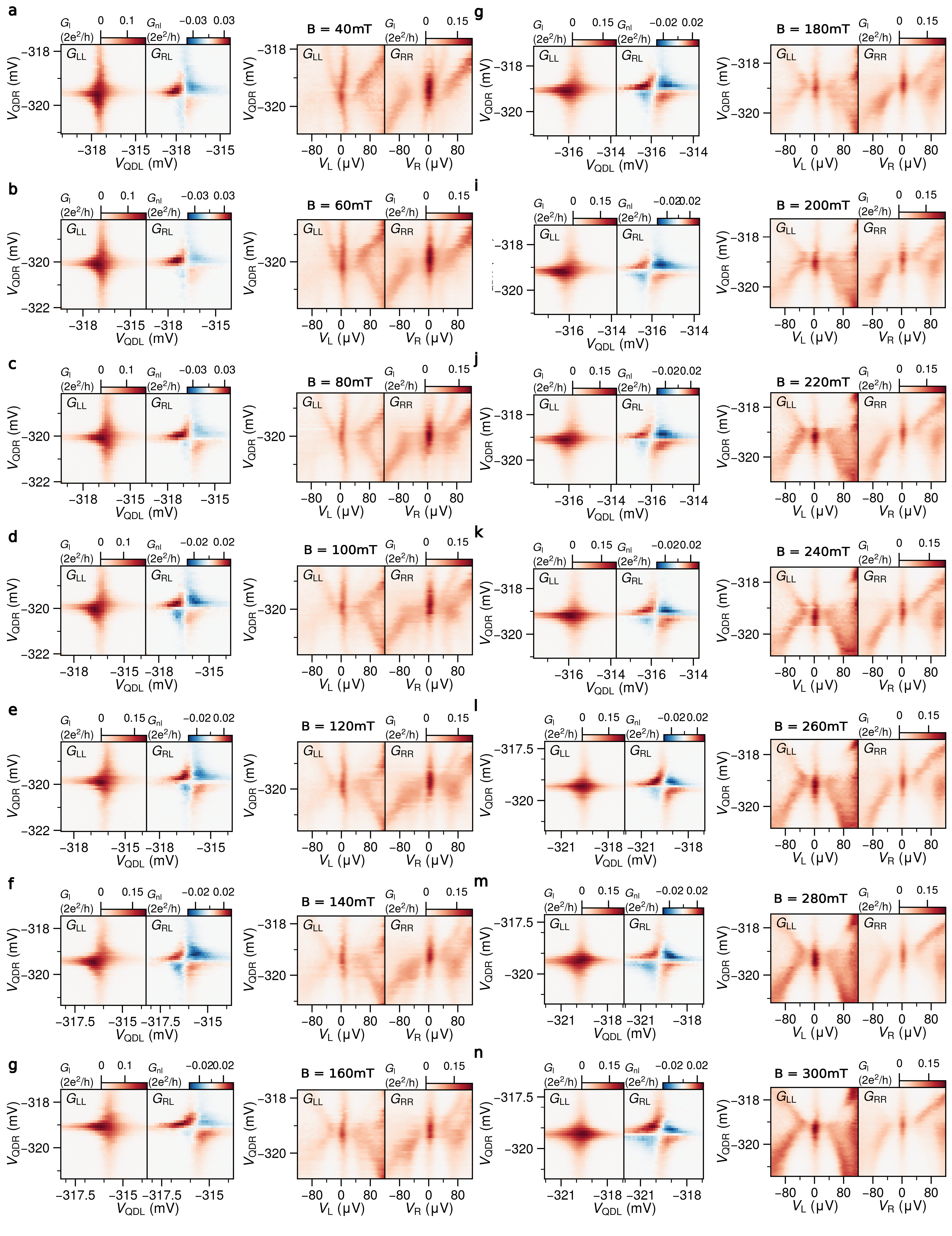}
\caption{\textbf{Raw datasets for Figure 4h.} 
Obtained 'sweet spots' at magnetic fields between \SI{0}{mT} and \SI{300}{mT}. \textbf{(a-i)} CSDs and tunnelling spectroscopy are measured at each sweet spot, where $V_{\mathrm{QDR}}$ is detuned. From these measurements $E_{\mathrm{det}}$ and $E_{\mathrm{gap}}$ are extracted, as described in the main text and in SI-Methods.}
\label{figS10}
\end{figure}

\begin{figure}[h!]
\centering
\includegraphics[width=0.95\textwidth]{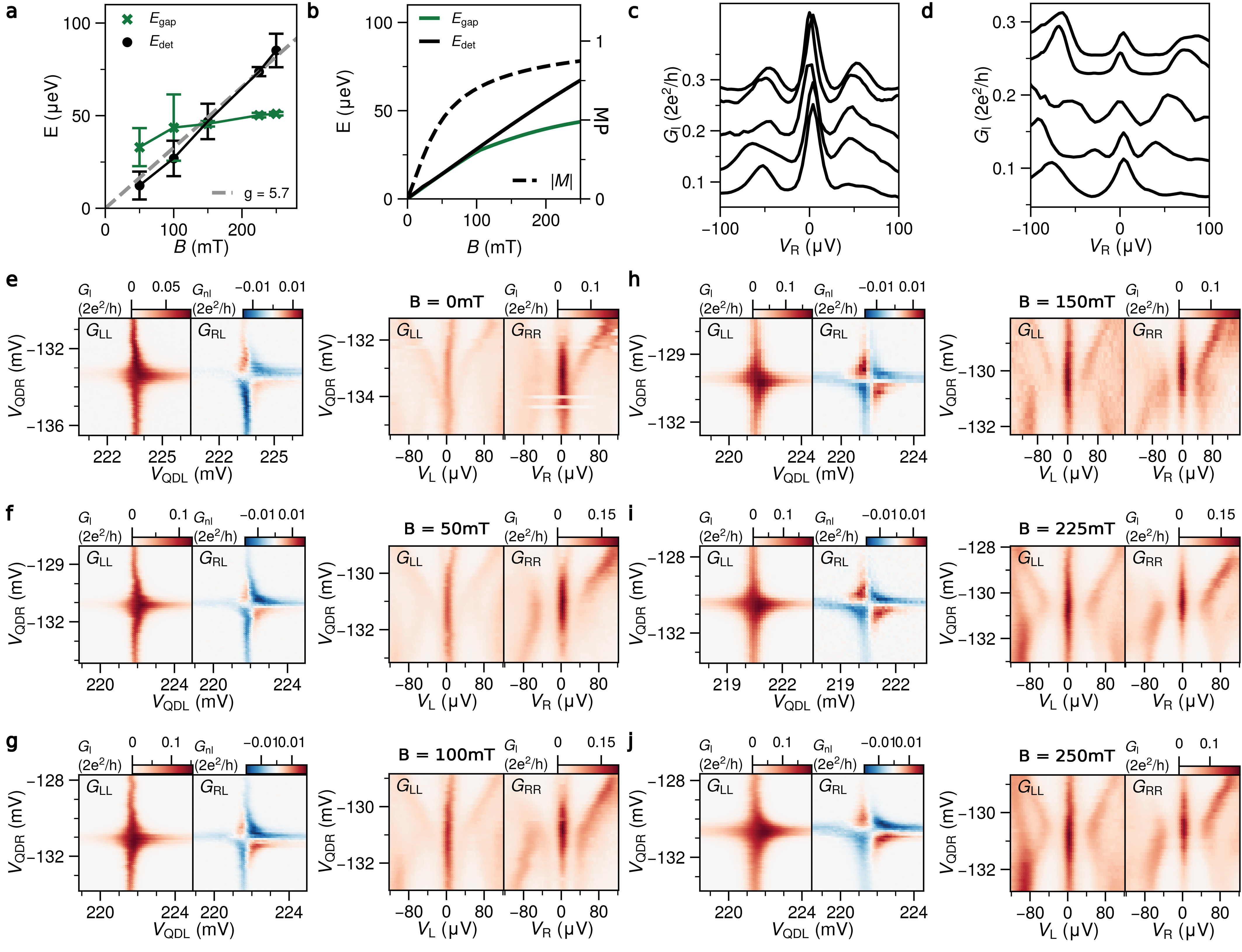}
\caption{\textbf{Extended datasets supporting Figure 4h.}
Reproduction of the main results from Fig.~4, using the orbitals shown in Fig.~3. 
Data was obtained at 6 different field values $B$ between 0 and \SI{250}{mT}. 
At each field $V_{\mathrm{ABS}}$ is adjusted to tune to the sweet spot. 
\textbf{a} Extraction of $E_{\mathrm{det}}$ and $E_{\mathrm{gap}}$, similar to the analysis presented in Fig.~4l. 
From a linear for of $E_{\mathrm{det}}$ a g-factor of 5.7 is estimated. 
\textbf{(b)} Numerically obtained $E_{\mathrm{det}}$ and $E_{\mathrm{gap}}$, using parameters tuned to compare to (a). 
At 250mT, an estimate of $M\approx 0.9$ is obtained. Extrapolation for comparison to Fig.~4l yields  $M\approx 0.92$ at 300mT.
\textbf{(c)} and \textbf{(d)} Waterfall plots highlighting the line-traces used to extract the data in (a). 
\textbf{(e-j)} Raw datasets of CSDs and tunnelling spectroscopy measurements, from which (a-d) is extracted. 
Datasets at \SI{150}{mT} and \SI{225}{mT} datasets are repeated from Fig.~3 and Fig.~S5 respectively.}
\label{figS9}
\end{figure}

\clearpage
\bibliography{ref} 

\end{widetext}
